\documentclass[aps,prb,twocolumn,superscriptaddress,showpacs]{revtex4-2}
\usepackage{graphicx}
\usepackage{bm,ulem}
\usepackage{amsmath,amsfonts}
\usepackage{amsbsy}
\usepackage[colorlinks=true,linkcolor=blue,urlcolor=blue,citecolor=blue]{hyperref}
\usepackage[utf8]{inputenc}
\DeclareUnicodeCharacter{03BC}{\mbox{$\mu$}}
\DeclareUnicodeCharacter{2009}{ }
\usepackage{color}
\usepackage{amssymb}
\usepackage[utf8]{inputenc}
\usepackage{colortbl}

\usepackage{xcolor}

\definecolor{imcolor}{rgb}{0.5,0.,0.5}

\newcommand{\ket}[1]{| #1 \rangle}
\newcommand{\bra}[1]{\langle #1 |}
\newcommand{\matrixel}[3]{\langle #1 | #2 | #3 \rangle}

\newcommand{\beq}{\begin{equation}}
\newcommand{\eeq}{\end{equation}}

\newcommand{\bea}{\begin{eqnarray}}
\newcommand{\eea}{\end{eqnarray}}
\newcommand{\lt}{<}

\newcommand{\dg}{\ensuremath{\dagger}}

\newcommand{\lr}[1]{\left( #1 \right)}

\begin{document}
%\title{Scar-like states at the quantum Hall edge}
\title{
Majorana braiding in superconductors with fixed total number of particles}
\author{Ivar Martin}
\affiliation{Materials Science Division, Argonne National Laboratory,
  Argonne, Illinois 60439, USA}
\author{Kartiek Agarwal}
\affiliation{Materials Science Division, Argonne National Laboratory,
  Argonne, Illinois 60439, USA}
\affiliation{Physics Department, McGill University, Montr\'eal, QC, Canada H3A 2T8}

\date{\today}

\begin{abstract}

One-dimensional  topological superconductors treated at the mean-field level host zero-energy edge Majorana modes, which encode topological degeneracy of their ground states. Geometric manipulations (braiding) of multiple wires can be used to induce topologically robust transformations within the ground state subspace.  
The mean-field ground states do not have a definite number of particles and thus cannot describe an isolated system. Projecting such states onto fixed particle number gives a very good approximation to the true ground state of an isolated superconductor. In previous work, we showed that so projecting the prototypical Kitaev wave function of a single wire retains key features of the mean-field description, such as the zero-energy single-particle spectral peaks near the wire edges. 
Here we consider the case of multiple wires with conserved total charge. Again, using the projected Kitaev wave functions as the template, we identify the  many-body counterparts of the mean-field topologically degenerate ground states. We then demonstrate how the quantum gates equivalent to braiding of Majorana zero modes can be implemented, albeit with reduced protection from noise and disorder.

%In the previous work  [SAM] using same approach we constructed MZM operators for a single wire and demonstrated the presence of a zero-energy peaks in the single-electron spectral function localized near the wire edges mimicing the mean-field MZMs. However, for a single wire with a fixed number of fermions there is no ground state degeneracy that would allow to perform nontrivial transformations within the ground state subspace. From the mean-field treatment, we expect that the topological degeneracy emerges when there are two or more wires (with four or more edge MZMs). Here, we identify the many-body states that correspond to this topological degeneracy. In the mean-field setting, unitary transformations within this ground state subspace  can be accomplished by braiding MZM's or by temporarily turning on interaction (tunneling) between MZM's. While our approach does not naturally allow geometric braiding,  we show that turning on tunneling between wire edges implements the same transformations in the ground state subspace as in the mean field case.

\end{abstract}

\maketitle

\section{introduction}

Protection of the quantum states from decoherence and the need for exquisitely precise quantum gates to control them represent two major obstacles towards realizing practical quantum computers. Topological quantum computing \cite{nayak2008non} has been touted as a novel approach to addressing these challenges---quantum information can be encoded non-locally in a topologically-protected degenerate ground state manifold, which is  immune to dephasing by local environmental noise, while certain quantum gates (potentially a universal set) can be realized in a topologically robust manner by braiding non-Abelian excitations. 

Some topological models can be implemented on top of the standard  superconducting qubit platforms \cite{bravyi2023high} or cold atom systems \cite{verresen2021prediction, bluvstein2022quantum}. However, it would be particularly appealing to realize topological states natively within material systems, which could allow achieving scalability more readily. An important example of such a topological system is the topological superconductor (TS).  In TSs, vortices (or surfaces) can contain Majorana fermion modes at zero energy (Majorana zero modes, or MZMs). These zero modes encode the ground state degeneracy, and the quantum states themselves can be manipulated by braiding MZMs around each other~\cite{ivanov2001non,mascot2023many}. 

While there are some bulk candidates for topological superconductivity, such as Sr$_2$RuO$_4$ or  UTe$_2$ \cite{mandal2023topological}, it can also be achieved at the interfaces between topological insulators and conventional superconductors \cite{PhysRevLett.100.096407}. Even more remarkably, one-dimensional (1D) TS wires can be realized by interfacing conventional superconductors and semiconductors with spin-orbit interaction \cite{PhysRevLett.105.077001, PhysRevLett.105.177002}. The latter scheme has attracted significant attention %\ks{as it appeared to require only a minimal amount of materials development}
as it involves interfacing well-understood materials~\cite{mourik2012signatures, lutchyn2018majorana}; yet, it has  proven to be an immense practical  challenge.

There are, however, also conceptual subtleties associated with small (quasi-) 1D superconductors; namely, for small systems the typical assumptions of the Bardeen-Cooper-Schrieffer (BCS) theory used to describe these systems are expected to fail. In particular, in small isolated superconductors, the number of electrons is fixed, in contrast to the usual grand-canonical BCS wave function, which breaks charge conservation symmetry.
Does the physics associated with MZMs survive beyond the BCS description \cite{lin2018towards}?  What are the special features of the true many-body states that encode the spectral characteristics of the localized MZM and their braiding properties?

In recent work~\cite{sajith2023signatures} we began answering these questions by studying the properties of Kitaev wave functions, which describe a topological superconductor in one dimension (1D), after it is projected on to a fixed number of electrons. These projected BCS states can be shown to be generally good approximate (variational) ground states of the number-conserving interacting Hamiltonians~\cite{leggett2006quantum}. The number-projected Kitaev wave function, in fact, can be shown to be the exact ground state(s) of 1D chains of spinless fermions with attraction \cite{ Iemini15, WangNumberConserving}. 
In Ref.~\cite{sajith2023signatures} we showed that the projected Kitaev wave function retains key features of its mean-field parent, such as the zero-energy single-particle spectral peaks near the wire edges. We also  constructed number-conserving analogs of the Majorana operators that switch between ground states whose particle numbers differ by one and found characteristic signatures typically associated with MZMs in the topological entanglement entropy. It is notable that the projected wave functions themselves lack any obvious features that would indicate the presence of edge MZMs, in contrast to the edge-localized mean-field Bogoliubov quasiparticles.

Here we extend our previous work to the case of multiple wires with conserved total charge. We are able to obtain and interpret the mean-field topological degeneracy, and identify the analogs of  MZM braiding purely in the language of many-body wave functions.
We find that while nontrivial transformations in the ground state manifold are possible, imposing the global charge conservation  leads to diminished topological protection.

This paper is organized as follows. In section \ref{sec:Kitaev} we summarize the properties of the mean-field and the number projected Kitaev wave functions. In section \ref{sec:split} we describe the evolution of the wave function when a single wire is  split into two by adiabatically removing internal bond(s). Comparison with the mean-field case allows identification of the correspondence between the many-body wave function with total number conservation, and mean-field states which only conserve global fermion parity. In section \ref{sec:op} we describe two ways to implement operations analogous to MZM braiding in the number conserving setting: one is ``Hamiltonian braiding" wherein tunneling between wire edges is turned on for a specific time that corresponds to a swap of MZM states in the mean field, and the second, ``pseudo-geometric" braiding, introduces by hand a Berry phase acquired by single-particle electronic operators when one of the two wires is physically rotated. In section \ref{sec:dephasing}, we discuss the quantum coherence issues associated with preparation of vacuum states of MZMs in the number-projected setting, and gate operations mimicking braiding. We contrast number-projected case and the mean-field, pointing out the higher degree of fragility of the former due to a higher degeneracy of the ground state manifold. However, we also find signs of topological robustness inherited from the mean field setting. In this section we present a number of numerical results, which illustrate our general arguments.
In section \ref{sec:Heis} we describe the mapping between the fermionic model whose ground states are the number-projected Kitaev wavefunctions and the Jordan-Wigner transformed spin model, which happens to simply be the Heisenberg model with global SU(2) symmetry. This further helps to understand both the ground-state degeneracy and the peculiar robustness of some of the states and operations on them.  We conclude the main text with discussion in section \ref{sec:discussion}. 
Some technical details and derivations are relegated to appendices.

\section{Kitaev wave function: mean-field and number-projected}\label{sec:Kitaev}

The mean-field Kitaev wave function encodes the unique properties of $p$-wave superconducting wires that make them such an appealing platform for topological quantum computing---namely, a topological ground-state degeneracy in the case of multiple wires, and the ability to induce  non-Abelian transformations in the ground state manifold by means of geometrical manipulation of the wires. In this section we first summarize the properties of the mean-field Kitaev wave function, as well as its number-projected descendants, which can be realized as the ground state of certain interacting (non-mean-field) models.

\subsection{Kitaev Hamiltonian and wavefunction}
For completeness we summarize here key facts about the Kitaev mean field model for $p$-wave spinless superconductor \cite{kitaev2001unpaired}. On a tight-binding chain of length $L$, it is described by the Hamiltonian
\begin{equation}
H_{MF} = -\sum_{j = 1}^{L - 1} t a_{j}^{\dg}a_{j + 1}  - \Delta a_{j}a_{j + 1} + h.c. + \mu a_{j}^{\dg}a_{j} \label{eq:HMF}, 
\end{equation}
where $a_{j},$ $a^{\dg}_{j},$  are fermionic annihilation and creation operators for the $j$-th site, $\Delta$ is the superconducting gap, $t$ is a hopping amplitude, $\mu$ is the chemical potential, and $L$ is the chain length.

The topological phase is realized for $|\mu|\lt 2t$; it is characterized by the appearance of Majorana  modes near the system's edges, which in the infinite $L$ limit manifest as the zero-energy peaks in the single electron spectral function. 

The appearance of the zero-energy fermionic mode leads to the degeneracy of the ground states with odd and even numbers of fermions. At the ``sweet spot" in the parameter space, $t = \Delta$ and $\mu = 0$, they have a particularly simple form
\beq
\ket{{\Psi_{o (e)}}}_L = \frac{1}{\sqrt{ L\choose N}}\sum_{\sum n_i = o (e)} {\lr {a_1^\dag}}^{n_1}{\lr {a_2^\dag}}^{n_2}\ldots{\lr {a_L^\dag}}^{n_L}\ket 0 \label{eq:Psi}
\eeq
where $n_{j} = 0,1$ is the occupation number of the $j^{th} $ site \cite{alicea2011non}.

In the case of a single wire, this parity degeneracy is not accessible due to the global parity conservation. However, in the case of two or more wires, multiple degenerate ground states within each total parity sector exist (e.g. for two wires with the total even fermion parity, there are two degenerate states with even-even and odd-odd subwire parities).

\subsection{Number-Projected Kitaev wave function}

The Kitaev wave functions have a definite parity, but an  indefinite number of electrons. Here we are interested however in the case when the superconducting wire is electrically isolated, which will force the number of electrons to be fixed. Restricting the mean-field wave function to a fixed number of electrons $N$ is expected to accurately capture the variational state of a fully interacting (number-conserving) Hamiltonian~\cite{leggett2006quantum}. 

In fact, it is possible to show that for  the number-conserving Hamiltonian~\cite{Iemini15,WangNumberConserving}
\begin{eqnarray}
        H = -t\sum_{i = 1}^{L - 1} a^{\dg}_{j}a_{j + 1} + a^{\dg}_{j + 1}a_{j} - n_{j} - n_{j + 1}
        + 2 n_{j} n_{j + 1},
\label{eq:HHazzard}
\end{eqnarray}
all states 
\beq
\ket{{N}}_L = \frac{1}{\sqrt{ L\choose N}}\sum_{n_1+\ldots+n_L = N} {\lr {a_1^\dag}}^{n_1}{\lr {a_2^\dag}}^{n_2}\ldots{\lr {a_L^\dag}}^{n_L}\ket 0 \label{eq:PsiN}
\eeq
are ground states (with zero energy).

%Consider now an isolated tight-binding chain (wire) of length $L$ with $N$ spinless fermions. $N$ can be anything from 0 to $L$.  Let us assume that the ground state for each $N$ is the number-projected version of the Kitaev wavefunction [alicea] at the ``sweet spot" of the model parameters that corresponds to the  perfectly localized Majorana edge modes,

As we show in \cite{sajith2023signatures}, these wave functions inherit several properties of their mean-field parents, including zero-energy peak in the spectral function near the wire edges. There is also Majorana-like operators that connect between ground states $\ket{{N}}$ and $\ket{{N\pm 1}}$.

For an isolated wire with a fixed number of fermions $N$ there is a unique ground state $\ket{{N}}_L$. However, in case of two (or more) disconnected wires with the conserved $total$ number of fermions, there is  now a macroscopic degeneracy that corresponds to arbitrary allocation of fermions between the wires, all with the same energy. This ``number" degeneracy replaces the parity degeneracy in the case of the mean-field Kitaev wave function. Nevertheless, as we show, two states within this manifold which in certain sense correspond to a $0$ and $\pi$ phase difference across the two subwires remain special and operators connecting these states exhibit certain residual topological robustness. The additional degeneracy, however, does make topological quantum computing more challenging than in the mean-field setting. 

The goal of this  paper is to explore in detail how the number-projected wave functions behave when, for instance, a single wire is being split into subwires and otherwise  manipulated to implement operations that correspond to MZM braiding.

\subsection{Heisenberg chain in hiding}
\label{sec:Heispre}
Upon applying the Jordan-Wigner transformation, $a_i^\dag = \sigma^+_i \Pi_{j<i}(- \sigma^z_j)$, with $\sigma^{x/y/z}_i$ the spin-$1/2$ Pauli operators on site $i$, we find that the number conserving model, Eq.~(\ref{eq:HHazzard})  turns into the spin-$1/2$ ferromagnetic Heisenberg chain
\begin{align}
   H & =-t \sum_{i = 1}^{L-1} \left[ \sigma^x_{i}\sigma^x_{i+1} + \sigma^y_{i}\sigma^y_{i+1} + \sigma^z_{i}\sigma^z_{i+1}-1\right]/2
   \label{eq:HDspin}
\end{align}
 The ground state wavefunction corresponds to the fully-polarized state with total spin $S = L/2$. It is $L+1$-fold degenerate, which correspond to the states $\ket{S, m}$ with different possible values of $m=\sum_i S^z_i = \sum_i \sigma^z_i/2$, or equivalently, all possible total fermion numbers, $N = L/2 + m$. Thus, the number degeneracy seen above is due to a hidden SU(2) symmetry which is made explicit in the spin model. As we show in Section \ref{sec:Heis}, this symmetry, or proximity to it are, helpful for interpreting many features associated with MZMs in the mean-field treatment.

%\ivar{Kartiek, the following para is mysterious. It refers to Majoranas, and pairs of spins forming spin 1, but without explaination. Lets drop it for now? and just say that in Sec we also explan what the Majorana states are in the magnetic language}
%Moreover, as we will see, this symmetry is also responsible in part for the robustness of the system when the chain is split to create new `MZMs' from the vacuum. 
%The fully polarized ground state is special in that {\it any} pair of spin-$1/2$s must be found in a spin-$1$ state to be consistent with the fact that the full chain is in a spin $S = L/2$ state. (Similar arguments can be made for multiple spin-$1/2$s forming ferromagnetic clusters themselves.) Thus, the ground state continues to remain at least an eigenstate of the Hamiltonian with any SU(2) symmetric perturbation. 

\section{Splitting the wire}
\label{sec:split}

In the standard Majorana wire architecture, manipulation  within the ground state manifold can be accomplished by splitting, rotating, and reconnecting wires. These manipulations correspond to various braidings of MZMs, which according to the mean-field theory are  located at the ends of wires. 

In this section we will focus on the structure of the number-conserving wave function as a single wire is split into two subwires by introducing a weak link.

\subsection{Uniform wire wave function in subwire basis}\label{sec:wire-subwire}

First we  note that the ground state of the pristine wire with fixed number of particles (\ref{eq:PsiN}) can be expressed in terms of the direct products of subwire ground state wave functions, defined on their respective Hilbert spaces. Denoting the ground states of the first subwire as $\ket{N_1}_{L_1}$ and the second wire as $\ket{N_2}_{L_2}$,

\beq
\label{eq:PsiN_12}
\ket{N}_{L} = \sum_{N_1+N_2 = N} \sqrt{\frac{{L_1\choose N_1}{L_2\choose N_2}}{{ L\choose N}}} \ket{N_1}_{L_1} \otimes \ket{N_2}_{L_2}
\eeq

The normalization follows form the identity $\sum_{N_1+N_2 = N}{{L_1\choose N_1}{L_2\choose N_2}} = {{ L\choose N}}$. In the limit when $L,L_1,L_2$ and $N,N_1,N_2$ are large, applying the Stirling formula, the amplitudes of the wave function scale as $\exp(-\delta N^2/4\bar N)$, where $\bar N =  p (1-p) L_1L_2/(L_1 + L_2)$ and $\delta N = N_1 - p L_1$, and $p = N/L$.  That is, they are strongly peaked at states where the filling factor on the two subwires coincide and match that of the complete wire. 

It is a special feature of the Kitaev ground state that it can be expressed via a linear superposition of subwire ground states. This is not immediately obvious since the total Hamiltonian $H$ is not a sum of subwire Hamiltonians, $H_i$ but rather $H = H_L + H_R + H_\text{inter}$, 
where $H_\text{inter}$ is the Hamiltonian of the bond connecting the subwires 1 and 2. However, for the special case when the condition $t_i = \Delta_i$ is satisfied on every bond (between sites $i,i+1$) and $\mu = 0$, the set of eigenstates is the same, uniquely labeled by the Majorana bond parities. In particular, that means that the eigenstates of two disconnected wires ($H_\text{inter} = 0$) are the same as for the perfectly uniform wire (see Appendix \ref{app:psiu} for details). In the former case, the eigenstates of the full system are obviously the tensor products of the eigenstates of the subwires, which upon projection into a fixed total number sector explains (\ref{eq:PsiN_12}).

\subsection{Wave function in the presence of a  weak link}
\label{sec:2W}

We now turn to the case when the link connecting subwires 1 and 2 is  weak but finite. This situation can be obtained starting from a uniform wire  and adiabatically decreasing $H_\text{inter}$. 
As noted above, at the mean-field level, if $t_i = \Delta_i$, the eigenstates remain unchanged with the changing strength of the bond. 
However, we will now approach the problem more generally, without relying on a specific form of the Kitaev Hamiltonian. 

Adiabatic theorem ensures that if the bond is weakened slowly enough, the system remains in the ground state. At a very weak $H_\text{inter}$ that means that the ground state is to a high accuracy a linear superposition of products of the ground states of disconnected subwires,  $\ket{N_1}_{L_1} \otimes \ket{N_2}_{L_2}$, which can be found via degenerate perturbation theory  with respect to $H_\text{inter}$ within that subspace. That is, it has the same form as \eqref{eq:PsiN_12}, but, generally, with a different set of amplitudes that depends on the specific $H_\text{inter}$.
For brevity, we can denote $\ket{N_1}_{L_1}\otimes\ket{N_2}_{L_2}\equiv\ket{N_1; N}$, since the total number of particles and the number of particles in the first subwire fully specify such states.

In this basis, the single electron tunneling between the subwires can be represented as 
\begin{equation}
    H_{\text{inter}} = - \sum_{n} t_{n} (\ket{n+ 1; N}\bra{n; N}+ h.c.). \label{eq:HT}
\end{equation}
Here, $t_{n}$  is a weak tunneling matrix element between subwires connecting states with $n$ and $n+1$ fermions in the first subwire. It is proportional to the hopping in $H_\text{inter}$. The tunneling $t_{n}$ is smooth positive function of $n$ due to the fact that tunneling matrix element depends on the wire fillings, but not on the exact electron numbers when the lengths of subwires are much larger than one and fillings are away from 0 and 1 (see Appendix \ref{app:tun} for explicit treatment of some special cases). 

The Hamiltonian \eqref{eq:HT} captures the low energy limit when we can neglect excitations outside the degenerate ground state manifold of individual wires.
A special feature of this Hamiltonian is that it represents coherent {\it single electron} tunneling (in contrast  to the Josephson Hamiltonian, which coherently tunnels Cooper {\it pairs}).
It corresponds to a tight-binding lattice with position-dependent hopping. Under the smoothness assumptions stated above, the ground state 
\begin{equation}
    \Psi_g = \sum_{n} c_{n}^g \ket{n; N} \label{eq:Psi0}
\end{equation}
has the amplitudes $\{c_{n}\}$  localized near the  maximum of 
$t_{n}$ (see Appendix \ref{app:wf} for details) with the energy being approximately $\epsilon_g = - 2 \max(t_n)$. 
%At half-filling ($p = 0.5$), the maximum is always reached at the natural filling, $n = L_1/2$, regardless of  $L_1/L_2$.

\begin{figure}
    \includegraphics[width=0.5\textwidth]{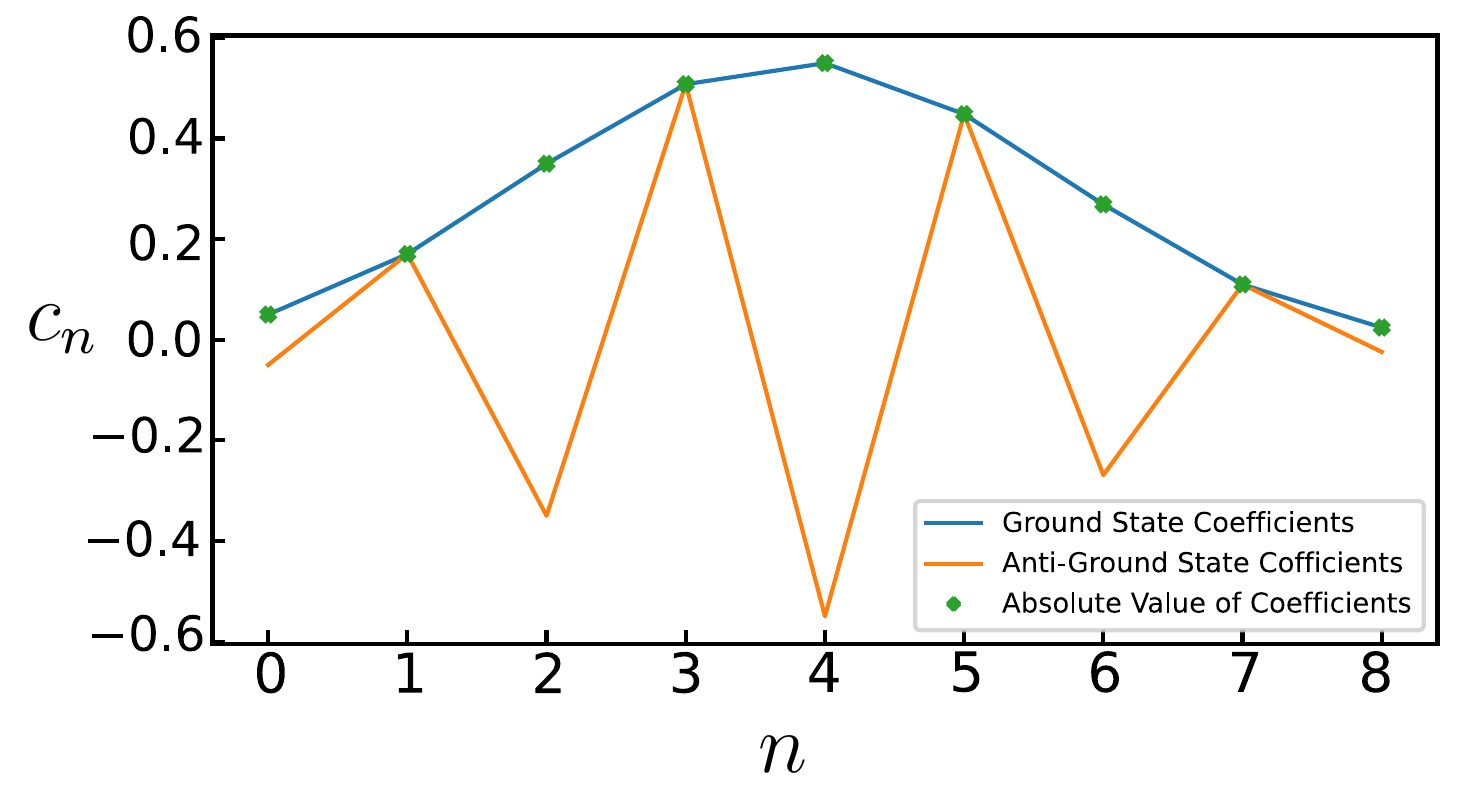}
    \caption{Coefficients $c_n$ of the ground state $\Psi_g$ and anti-ground state $\Psi_{ag}$ of the split wire in the basis of eigenstates $\ket{n; N}$ of $H_L + H_R$. For details of the computation, see main text under Eq.~(\ref{eq:Hinter})}
    \label{fig:coeffs}
\end{figure}

By the Perron-Frobenius theorem, all $c_{n}^g$ are non-negative.
From the symmetry of the Schr\"odinger equation it also immediately follows that the highest energy (anti-ground) state is 
\begin{equation}
    \Psi_{ag} = \sum_{n} c_{n}^{ag} \ket{n; N} = \sum_{n} (-1)^n c_{n}^g \ket{n; N}, \label{eq:Psi1}
\end{equation}
with the approximate energy  $ \epsilon_{ag} = 2 \max(t_{n})$. We verify these expectations numerically using fully interacting model of a chain \eqref{eq:HHazzard} with a weak link; see Fig.~\ref{fig:coeffs}. Here we computed the number projected states by exactly diagonalizing the Hamiltonian

\begin{align}
    H &= H_L + H_R + H_{\text{inter}} \nonumber \\
    H_{\text{inter}} &= - t_{\text{inter}} \left( a^\dagger_{L_1} a_{L_1 + 1} + a^\dagger_{L_1 + 1} a_{L_1} \right) 
    \label{eq:Hinter}
\end{align}
for a very small value for $t_{\text{inter}} = 10^{-8} t$, and projecting the ground state obtained on to fixed number states in the subwires. $H$ was again diagonalized for $t_{\text{inter}} = 10^{-3}t $ and the ground and anti-ground states of this Hamiltonian were designated to be $\Psi_g, \Psi_{ag}$ from which overlaps on the number projected states were computed. We set $t = 1$ without loss of generality. The anti-ground state was chosen to be the largest energy state in the ground state manifold which is identifiable due to its finite energy separation from all other  excited states. 

These two states happen to correspond precisely to the  special mean-field states with the even and odd parity of the Majorana fermions  at the tunnel junction.  To demonstrate that, let us compare the weak bond limit in the mean field and in the many-body cases. 

In the mean field Kitaev Hamiltonian,  when the tunneling between two  subwires is adiabatically turned off completely,  there are two new MZMs created next to the broken bond. The parity of this pair is taken as even, since it was created from the ``vacuum" (bulk of a perfect wire). Adiabatically changing $H_\text{inter}$ to the original value will restore the uniform state of the intact wire with no qusiparticles created.  

On the other hand, starting from the ground state of an interacting Hamiltonian \eqref{eq:HHazzard} with a fixed number of fermions $\ket{{N}}_L$,  and gradually reducing the tunneling between subwires produces the state $\Psi_g$ of Eq. (\ref{eq:Psi0}). This  suggests that  this state is the fixed-number counterpart of the mean-field state with even Majarana parity at the junction. 
Further, let's consider the scaling of the energy of the $\Psi_{g/ag}$ states with the strength of the interwire tunneling. 
In the mean field treatment, the energy of two MZMs  coupled at the junction by the local electron tunneling $t_{\text{inter}}$ is given by $-i t_{\text{inter}}\gamma_2\gamma_3$ (see Fig.~\ref{fig:schematic}); that is,  the energy is proportional to the tunneling matrix element and the parity  $i \gamma_2\gamma_3$ of the MZM pair in the middle of the partially split wire. On the other hand, the matrix elements in the effective lattice Hamiltonian \eqref{eq:HT} are also proportional to the strength on the local electron tunneling, yielding the energies of the ground and anti-ground states as $\pm 2\max t_n$, where $t_n \propto t_{\text{inter}}$. We thus are led to the conclude that the ground state in the number conserving case corresponds to the even Majorana parity in the mean field,  $i \gamma_1\gamma_2 = +1$, and the anti-ground state -- to the odd parity state $i \gamma_1\gamma_2 = -1$. 

It is interesting to note that it is possible to convert between $\Psi_g$ and $\Psi_{ag}$ by applying a potential pulse to one of the subwires---this temporarily shifts the chemical potential, making the phase wind in proportion to the number of fermions in the subwire. That is, $\ket{n; N} \to e^{in\phi}\ket{n; N}$, where $\phi$ is the time integral of the potential pulse. This effect is 
nontrivial since  both $\Psi_g$ and $\Psi_{ag}$ are superpositions of different number states. In particular, the application of phase $\phi = \pi$  (``$\pi$-pulse") converts between the states $\Psi_g$ and $\Psi_{ag}$. 

\subsection{Wire parity basis }

The many-body states $\Psi_g$ and $\Psi_{ag}$, as we argued above, correspond to the opposite parity states  of the MZM pair at the tunnel junction.  The even Majorana parity state is the ground state; the fact that different number states enter with the same phase in the superposition  is analogous to having zero superconducting phase difference between the two sides of the junction in the mean-field. The odd parity state corresponds then to the single-fermion phase difference $\pi$ between the superconductors in the mean field language. 
%\footnote{The analogy with the mean field  is not precise, since for a finite number of fermions that we consider the conjugate phase can only be defined approximately.
%Furthermore, the phase in conventional superconductors is the phase of the Cooper pairs; hence it is twice our phase $\phi$, meaning that $\phi = \pi$ is equivalent to the zero phase difference for the Cooper pairs across the junction.}  

Another important  basis can be obtained by combining these two states as 
 \begin{eqnarray}
     \Psi_e = \frac{\Psi_g  + \Psi_{ag} }{\sqrt{2}} = \sqrt{2} \sum_{n - even} c_{n}^g \ket{n; N}\label{eq:even}\\
     \Psi_o = \frac{\Psi_g  - \Psi_{ag} }{\sqrt{2}} = \sqrt{2} \sum_{n - odd} c_{n}^g \ket{n; N}
 \end{eqnarray}
In contrast to the junction basis, this is the $wire$ parity basis, which has either even or odd numbers of fermions in the first subwire (in the mean field, this translates into the even or odd parity of the pair of edge MZMs that belong to the same subwire). 

We finally note that unlike $\Psi_{g/ag}$, the
states  that correspond to the ``phase differences" other than 0 and $\pi$ do not yield exact eigenstates of the tunneling Hamiltonian \eqref{eq:HT}, see Appendix \ref{app:approx} for details. Yet the expectation values of their energies and junction currents follow the expected Josephson-like form.

%It is interesting to note the analogy between these states and the bosonic cat states (which are the superpositions over even and odd boson mode occupation numbers).

%In Appendix \ref{app:approx}, we  discuss other  eigenstates of this system, which approximately correspond to the phase differences on the junction other than 0 and $\pi$. Of all these states, $\Psi_g $ and $\Psi_{ag}$ are maximally separated from each other in the space of $\phi$. In the meantime, a typical noise that would affect this system is a gate voltage (equivalent to the chemical potential) noise. Such noise introduces drift in phase $\phi$, and if not too large, should be possible  to correct for, bringing the system back to the computational subspace.

\section{Quantum operations}
\label{sec:op}

\begin{figure}
    \includegraphics[width=0.43\textwidth]{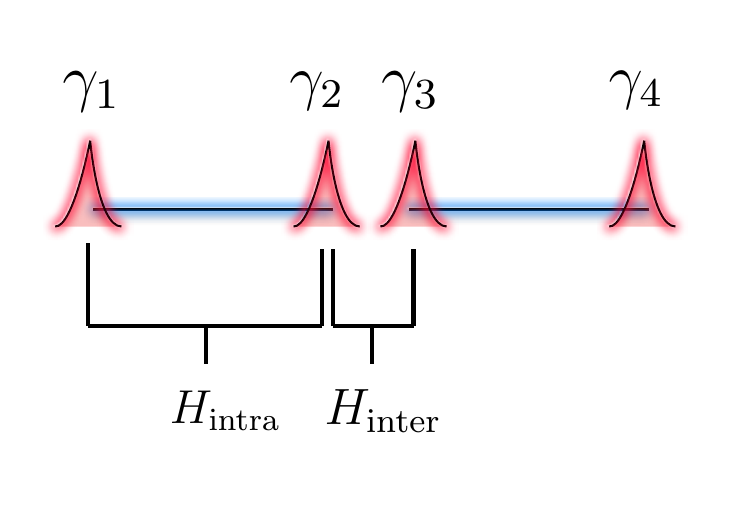}
    \caption{Schematic figure of the superconducting one-dimensional wire split into two subwires with MZMs $\gamma_1,\gamma_2, \gamma_3, \gamma_4$ at the ends (in the mean-field setting). Also shown are tunnel couplings between ends of the left subwire labeled $H_{\text{intra}}$, and the terms coupling the two wires, $H_{\text{inter}}$.}
    \label{fig:schematic}
\end{figure}

Simple adiabatic splitting and reconnection process of a superconducting wire should leave the state  $\ket{{N}}_L$ intact. However, if an additional energetic or geometric manipulation is performed along the way, the final state can differ from the initial. This is conveniently interpreted at the mean-field level in terms of manipulations of MZMs (by braiding or tunnel coupling).

We will now examine the corresponding effects in the number conserving setting. Above we already saw that splitting a single wire into two subwires, applying relative potential pulse $\pi$ and rejoining the subwires, instead of returning to the ground state, takes it to the anti-ground state with the energy $4 \max t_n$ above the ground state. 

%\textcolor{red}{This para could probably be done without in this location?} After fully patching the wire,  this excess energy (of the same order as the mean-field superconducting gap in the Kitaev  model), will determine the excitation level of the final state (in the mean-field it corresponds to creation of quasiparticles above the ground state). Detection of such excitation can be used  to distinguish between the  $\Psi_g$ and $\Psi_{ag}$ states.

In the following we will consider various control protocols that can be applied to the number-projected wire states to induce transformations in the many-body ground state manifold, and relate them to the corresponding  MZM  manipulations in the mean field.

%Now that we identified the relevant  basis states, we can ask how these many-body states transform under operations that correpond to MZM braiding. 

In the T-junction architecture \cite{alicea2011non}, braiding is accomplished by spatially rotating, fusing and splitting quantum wires.  While splitting and fusing the wires has to do with the chain topology, and therefore can be managed within the number-conserving model here, rotations also involve geometry (state dependence on the wire orientation relative to the substrate/underlying crystal lattice), which pure 1D models lack. For instance, in Alicea et al \cite{alicea2011non},  the topological superconducting wires have globally-imposed orientation-dependent superconducting phase (e.g., the order parameter phase difference between horizontal and vertical wire segments is $\pi/2$). 

Here we propose two ways to capture the effects of geometry in the number-conserving many-body setting: (1) by including the effects of wire rotation on the single-electron wave functions, and hence also on the many-body wave function; or (2) by replicating the MZM braiding by means of Hamiltonian evolution, which simply corresponds to coupling the ends of the subwire(s) for a precise length of time that is equivalent to their braiding in the mean-field picture. 

In this section, after refreshing the basics of the MZM braiding in the mean field, we will demonstrate how it can be implemented in this number conserving scheme.

\subsection{Mean-field MZM braiding}
\label{subs:MFb}
A representative example of MZM braiding in mean field starts with two subwires with well-defined parity each, and exchanging the MZMs that belong to different wires.  This can be done using the T-junction geometry of Alicea et al \cite{alicea2011non}. The first wire contains MZMs $\gamma_{1,2}$ at the edges and the second contains  $\gamma_{3,4}$; see Fig.~\ref{fig:schematic}.

The exchange of MZMs can be described by the unitary operation \cite{ivanov2001non}
\begin{equation}
    T_{23} = \exp{\frac\pi 4\gamma_2\gamma_3},\label{eq:T}
\end{equation}
which converts $\gamma_2 \to \gamma_3$ and $\gamma_3 \to -\gamma_2$. The effect of this transformation is generally nontrivial in the multidimensional ground-state manifold, whose basis states can be labeled by the parities of the constituent MZM pairs. Notably, it can be accomplished  by moving MZM around each other in real 2D space, with the result depending only on topology of the exchanges (braiding) but not the geometry of the MZM paths. This is the essence of the topological quantum computing using MZMs.

Under single, e.g. clockwise, exchange of $\gamma_2$ and $\gamma_3$ the new state becomes a superposition of the wire parity states.  For example, 
\begin{eqnarray}
    \Psi_{ee} \to \frac{\Psi_{ee} + i \Psi_{oo}}{\sqrt{2}},\label{eq:mfbr}\\
    \Psi_{oo} \to \frac{\Psi_{oo} + i \Psi_{ee}}{\sqrt{2}}.
\end{eqnarray}
Here $\Psi_{ee/oo}$ are the  mean-field ground states of two disconnected wires. Here the first (second) letter of the subscript marks the parity of  the first (second) subwire.  We use two labels here as a way to distinguish these wave functions with indefinite total number of fermions from the many-body wave functions above that have a fixed total number of fermions.

The second exchange in the same direction leads to $\Psi_{ee} \to\to  [{(\Psi_{ee} + i \Psi_{oo}) + i (\Psi_{oo} + i \Psi_{ee})}]/{{2}} = i \Psi_{oo}$, i.e. a perfect parity switch. 

An equivalent effect can be obtained by  starting from the definite parity state of the junction MZMs (e.g. $i\gamma_2\gamma_3 =1$) and exchanging MZMs that belong to the same wire,  e.g. $\gamma_1$ and $\gamma_2$.  This variant can be implemented by starting from a single wire, adiabatically splitting it into two, followed by rotation of the first wire that replaces $\gamma_2$ with $\gamma_1$ at the junction. As a result, the state will become a superposition of junction parity states of the same kind as just described.

From (\ref{eq:T}) it is clear that the MZM brading transformation can be also  implemented in a non-topological fashion, by turning on the hopping between $\gamma_2$ and $\gamma_3$, i.e. tunneling between the wires, for the period of time that corresponds to the phase accumulation $\pi/4$. The same operation can be applied to any other pair of MZMs, whether they belong to the same or different wires.
We will call this approach {\it Hamiltonian braiding.} 

Next we will see how the braiding can be implemented in the number-conserving setting.

\subsection{Hamiltonian braiding}
\label{sec:HamiBraid}

In the mean field approach, the unitary (\ref{eq:T}) can be implemented by introducing tunneling $t_{\text{inter}}$ between two ends of the subwires, where the MZMs $\gamma_2$ and $\gamma_3$ reside. 
As long as $t_{\text{inter}}$ is smaller than the bulk spectral gap, at the mean field level, $t_{\text{inter}}$ only affects the MZMs (not the bulk/quasiparticle states). When turned on for a specific duration of time $\tau$, such that $t_{\text{inter}} \tau = \pi/4$, this matrix element can, in particular,  implement single exchange of MZMs. Other tunneling times  in general lead to non-quantized transformations.

In the many-body approach, the wave function of two weakly coupled wires with fixed total number of electrons evolves according to   the tunneling Hamiltonian, Eqs.~(\ref{eq:HT}, \ref{eq:Hinter}), 
\begin{eqnarray}
    i\frac{d}{d\tau} \Psi = \left( H_L + H_R + H_{\text{inter}} \right) \Psi.
\end{eqnarray}
To illustrate the Hamiltonian braiding in this case, let us suppose that the initial state  corresponds to the even many-body wire parity state  $\Psi_e$, Eq. \eqref{eq:even}. 
The initial state can be expanded into the eigenstates of the tunneling Hamiltonian as
\begin{eqnarray}
    \Psi(0) = \Psi_e = \frac{\Psi_g + \Psi_{ag}}{\sqrt{2}}.
\end{eqnarray}
The time evolution is now easy to compute, with 
\begin{eqnarray}
    \Psi(\tau) =  \frac{\Psi_g e^{i\epsilon_{g}  \tau}+ \Psi_{ag} e^{-i\epsilon_{ag}  \tau}}{\sqrt{2}}. \label{eq:hbraid1}
\end{eqnarray}
If we now choose $ \epsilon_{ag}\tau = -\epsilon_{g}\tau =\pi/4$, then
\begin{eqnarray}
    \Psi(\pi/4) =  \frac{\Psi_g + \Psi_{ag} + i  (\Psi_g - \Psi_{ag})}{{2}} = \frac{\Psi_e + i  \Psi_o }{{\sqrt{2}}}.
\end{eqnarray}
And naturally, at twice the interaction time 
\begin{eqnarray}
    \Psi(\pi/2) =  i  \Psi_o.
\end{eqnarray}
These results are in complete agreement with the expectations for MZM braiding at the mean field level, Eq. \eqref{eq:mfbr}, since the energies $\epsilon_{g/ag}$ are obviously proportional to $t_{\text{inter}}$. This further supports the identification of $\Psi_{g/ag}$  and $\Psi_{o/e}$ as the many-body counterparts of the Majorana parity states in the junction and the wire bases, respectively. 

It should be noted that the key ingredient is the phase coherence between states in the superposition that correspond to different numbers of electrons in the left and the right subwires. This coherence is possible even when the combined number of electrons in both wires is fixed.

Similarly, from the mean-filed  we expect that exchanging MZMs $within$ one subwire  should have a nontrivial effect on the interwire junction parity states. We can implement this braiding by turning on  tunneling between the first and the last sites of the first subwire,  
\bea
H_{\text{intra}} = -t_{\text{intra}} (a_1^\dag a_{L_1} + a_{L_1}^\dag a_{1}). 
\label{eq:HTintra}
\eea
Let us study the effect of such tunneling on the many-body states $\Psi_{g/ag}$, which we expect to correspond to the different junction MZM parities.

Restricting to the ground state manifold of two-wire states, this Hamiltonian leaves individual $\ket{n; N}$ states intact; however, they acquire non-zero energies $2 t_{\text{intra}} (-1)^{n} p_1(1-p_1)\equiv (-1)^{n}\tilde t_{\text{intra}}$, where $p_1= n/L_1$ \cite{sajith2023signatures}. 
Since the relevant number states are concentrated in a narrow window of $\delta p_1\sim 1/\sqrt{L_1}$, all the eigenvalues have essentially the same magnitude, but opposite sign for
even and odd $n$. 
The sign factor $(-1)^{n}$ is very important. It originates from the Fermi statistics: When the total number of electrons in the first subwire is odd, tunneling an electron from the first to the last site equivalent to an even number of permutations, and hence does not affect the sign of the wave function. In contrast, for even $n$, tunneling produces a ground state with the opposite sign. 
Therefore, the odd and the even wire parity states $\Psi_e$ and $\Psi_o$ will respond differently to the intra-wire tunneling. For instance, if we start from 
\begin{eqnarray}
    \Psi(0) = \Psi_g = \frac{\Psi_o + \Psi_{e}}{\sqrt{2}}
\end{eqnarray}
and turn on the intra-wire coupling, the evolution becomes
\begin{eqnarray}
    \Psi(\tau) = \frac{\Psi_o e^{i\tilde t_{\text{intra}}\tau}+ \Psi_{e}e^{-i\tilde t_{\text{intra}}\tau}}{\sqrt{2}}.
\end{eqnarray}
Following the identical reasoning as in the case of inter-wire tunneling, we conclude that turning on tunneling between the edges of the same wire is equivalent to ``braiding" intra-wire MZMs; that is, it produces the corresponding transformation in the ground state manifold. 

We therefore conclude that we have correctly identified the many-body counterparts of the mean-field MZMs parity states, which follow the same transformation rules as the  mean-field states under the ``Hamiltonian braiding." 

\subsection{Pseudogeometric braiding}
The standard way to implement MZM braiding in the quasi-1D setting involves rotation of the  superconducting wires with respect to a static background \cite{alicea2011non}. The orientation-dependence of the mean-field superconducting order parameter then leads to the desired transformations in the space of degenerate ground states. 

The number-conserving models, such as Eq. (\ref{eq:HHazzard}), despite sharing some features with the Kitaev mean-field model, do not have built-in $U(1)$-breaking superconducting order parameter. In fact, any projection of the Kitaev ground state to a fixed number of particles is a ground state of (\ref{eq:HHazzard}), while the mean field Kitaev model only has  two ground states distinguished by the fermion parity (the Hamiltonian mixes the number sectors, but preserves parity).  The absence of the rigid angular-dependent superconducting order parameter deprives the  number-projected wave function of any angular dependence. We will show however that by imposing angular dependence at the single-electron orbital level leads to the recovery of the desired mean-field braiding results. Since strictly speaking such angular dependence is extraneous to the interacting models of the kind (\ref{eq:HHazzard}), we brand this {\it pseudo}geometric braiding.

We will consider the following braiding protocol: (1) Start with a single wire with a specific number of fermions, (2) Split it into two subwires by weakening one of the bonds, (3) ``Rotate"  one of the subwires, which reverses the order of fermions in its wave function and imprints single-orbital Berry phase on electrons  (4) Compare the full system wave functions before and after the rotation. The result should be referenced to the mean field expectation described in section \ref{subs:MFb}. 

%(1) Start with single wire with fixed fermion parity encoded by the two edge MZMs  $\gamma_1$ and $\gamma_4$, (2) Split the wire, generating two more MZM's, $\gamma_2$ and $\gamma_3$, with the trivial parity $i\gamma_2\gamma_3 = +1$, (since the pair is pulled out of ``vacuum"); the resulting wave function will be denoted  $\Psi_+$ indicating the positive junction parity. (3) Rotate  the first wire, which swaps MZM's $\gamma_1$ and $\gamma_2$ according to $\gamma_1 \to \gamma_2$ and $\gamma_2 \to - \gamma_1$. The mean-field wave function after single rotation  becomes $(\Psi_+ + i \Psi_-)/\sqrt{2}$, where $\Psi_-$ denotes the state where the parity of the MZMs at the junction is negative (the sign in front of $i$ depends on the direction of rotation) [Dirk].

As discussed above, adiabatically splitting the wire with fixed number of electrons leads to a ground state  $\Psi_g $, (\ref{eq:Psi0}).
Rotation of the first subwire of length $L_1$ reverses the order of sites 1 through $L_1$ in the wave function. To revert to the original basis states, we can reverse the order of fermions, keeping track of the permutations; this requires $n(n-1)/2$ fermion exchanges. Thus, after the ``rotation" the state becomes 
\begin{equation}
    \Psi' = \sum_{n}  (-1)^{\frac{n(n - 1)}{2}} c_{n} \ket{n; N}.
\end{equation}
It has the signature $++--++...$, which is distinct from the signatures $++++...$ for the ground state (\ref{eq:Psi1}), and $+-+-...$ for the anti-ground state (\ref{eq:Psi1}). It is related however to the signature of certain  approximate eigenstates $\Psi_{\phi}$, which  we discuss in Appendix \ref{app:approx}. 
 Indeed, 
\beq
\Psi' = (\Psi_{\pi/2} + i\Psi_{-\pi/2})/\sqrt{2i}.\label{eq:Psipr}
\eeq
While this is a nontrivial transformation in the approximate ground state manifold, it is  clearly different from the one expected from the braiding of the mean-field MZMs due to the lack of geometry -- a permutation is not the same as rotation.

An obvious deficiency of treating the  reversal of the fermion order as ``rotation" is that there is no sense of direction that would allow to distinguish clockwise form counter-clockwise rotations. It is also easy to verify that applying the procedure (\ref{eq:Psipr}) twice simply returns the state to the original $\Psi_g$ instead of producing the  parity switch to $\Psi_{ag}$ expected from the mean field. %\footnote{Note however that if the transformation did not mix positive and negative quasimomenta $\phi$, then, two exchanges in the same direction would transform state $\Psi_0$ into $\Psi_\pi$, which is the same answer that we would get in the mean field.}

In order to reintroduce geometry and thus recover the Majorana-like braiding properties  we impose nontrivial  transformation rules for individual fermion orbitals under rotation. For inspiration, we can take the case of a topological superconductor wire on the surface of a topological insulator. The electronic states in this case are characterized by the momentum-(pseudo)spin locking. Therefore, rotation of the wire by $\pi$, which reverses momenta of quasiparticles, results in the Berry phase $ \pi/2$ per electron (or $-\pi/2$ for the opposite direction of rotation). For a many-body wave function with $n$ fermions the extra phase factor is $e^{i\pi n/2}$.
Including this phase factor in $\Psi'$, Eq. (\ref{eq:Psipr}), we obtain
\bea
\Psi_+'' &=& \sum_{n} e^{ \frac{i\pi n}{2}} e^{\frac{i\pi n(n - 1)}{2}} c_{n} \ket{n; N} = \frac{\Psi_{\pi} + i\Psi_{0}}{\sqrt{2i}}.
\eea
For the opposite direction rotation we obtain $\Psi_-'' = ({\Psi_{0} + i\Psi_{\pi}})/{\sqrt{2i}}$. 
This result, $\Psi_\pm''\propto \Psi_{0} \mp i\Psi_{\pi}$ is precisely what we would obtain from braiding MZMs in the mean field [recall that $\Psi_{0}$ is the even junction parity state ($\Psi_g$), and $\Psi_{\pi}$ is the odd ($\Psi_{ag}$)]. 
Apart from establishing the connection between the number-conserving and the mean-field treatments, such orbital phase imprinting may also be a useful tool to implement braiding in numerical simulations.
\begin{figure}
    \includegraphics[width=0.43\textwidth]{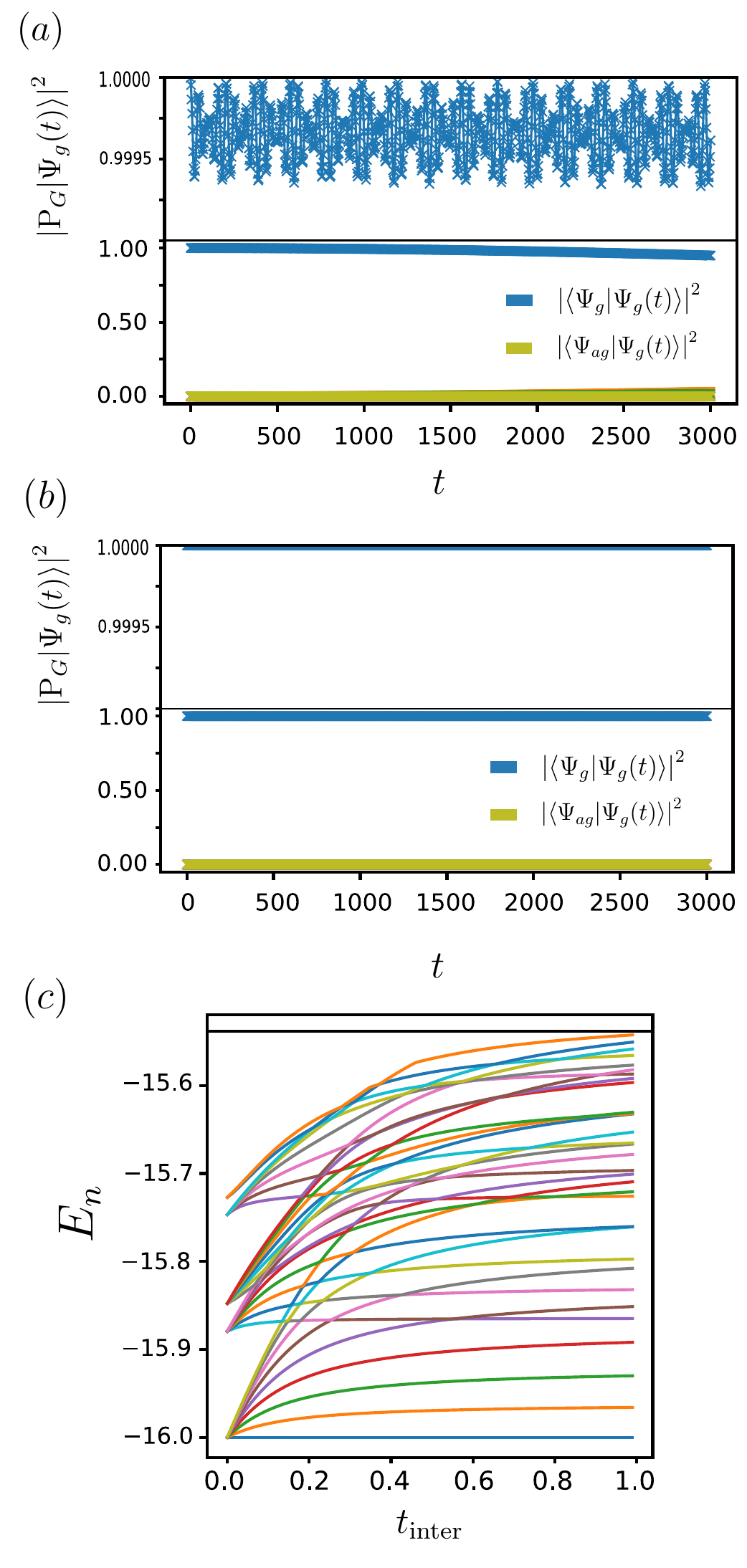}
    \caption{(a): Sudden quench from the ground state of the uniform wire as outlined in Eq.~(\ref{eq:quench}) and below. The system remains within the ground state manifold (top subfigure) with the overlap of the time evolved state on the ground state manifold remaining close to $1$. Within the ground state manifold (bottom subfigure), one observes small changes to the occupation of different states at long times; these can be suppressed in an adiabatic quench. (b): Sudden quench with an SU(2)-preserving change of the bond Hamiltonian to $H_b(0^+) = (t_{\text{inter}}/t) H_b (0^-)$ leads to no excitation of the system to numerical accuracy within the ground state subspace or outside of it.(c): Evolution of the eigenspectrum as the uniform wire, with $t_{\text{inter}} =1$, is split uniformly (decreasing proportionately the chemical potential, interaction strength, and hopping across the middle bond) into two decoupled wires at $t_{\text{inter}} = 0$. }
    \label{fig:quench}
\end{figure}

\section{Robustness of manipulations}
\label{sec:dephasing}

The main attraction of topological quantum computing  is  its robustness to environmental noises, which cause relaxation and dephasing, and robustness in the implementation of quantum gates.  In this section we perform a number of numerical calculations to test the validity of the general arguments presented above.

In the architecture of Ref. \cite{alicea2011non}, which assumes a global superconducting gate that pins the superconducting phase, the relevant quantum states are  fully specified by the parities of pairs of MZMs, and the excited states are separated from them by a finite energy gap.  
Then, the quantum operations induced by braiding of MZMs are expected to be topologically robust; that is, an operation only fails if the braiding operation is incomplete, or, for instance, a stray quasiparticle takes the system out of the computational subspace. 

In the number conserving case, the level of quantum protection can be expected to be lower for two reasons---i) there is a macroscopic $\mathcal{O} \left( L \right)$ ground state degeneracy of ground states when the wire is cut as opposed to merely two in the mean-field setting, and ii) the system is not gapped; there are  neutral excitations whose energy scales as $1/L^2$ in the symmetric point of (\ref{eq:HHazzard}), which we primarily focus on here, and $1/L$ for general parameters  (see \ref{sec:Heis}). However, we will now demonstrate that the level of protection is in fact substantially higher than what the above issues would naively lead us to expect. In particular, the two states $\Psi_{g}, \Psi_{ag}$ which are the counterparts of the MZM mean-field states, are indeed special. We will consider next the preparation of such states and their manipulation and discuss the robustness of these operations using numerical and analytical arguments. 

%Among the reasons are the much larger ground state degeneracy and the exposure to the decoherence by the charging energy. \kc{However, }

\subsection{State preparation}

When the Kitaev wire is split in half, one expects that two newly formed MZMs in the center are  in the `vacuum' state. In the number conserving case, this corresponds to the preparation of $\Psi_g$ starting from the ground state of the uniform wire. Specifically, we consider a time-dependent Hamiltonian that transforms, as ideally as possible, from the ground state of the uniform wire to $\Psi_g$---

\begin{align}
    H_{\text{prep}}&(\tau) =  H_L + H_R + H_b(\tau) + 2 \mu N \nonumber \\ 
    H_L = -&\sum_{i=1}^{L_1-1} t a^{\dagger}_{i} a_{i+1} + t a^{\dagger}_{i+1} a_{i} + \mu n_i + \mu n_{i+1} + U n_i n_{i+1}, \nonumber \\
    H_R = -&\sum_{i=L_1+1}^{L_2-1} t a^{\dagger}_{i} a_{i+1} + t a^{\dagger}_{i+1} a_{i} + \mu n_i + \mu n_{i+1}+ U n_i n_{i+1}, \nonumber \\
    H_{b} (\tau) &= -t_{b} (\tau) \left[ a^\dagger_{L_1+1} a_{L_1} + a^\dagger_{L_1} a_{L_1+1} \right] + U_{b} (\tau) n_{L_1} n_{L_1 + 1} \nonumber \\
    & - \mu_b (\tau) n_{L_1} - \mu_b (\tau) n_{L_1 + 1}.
    \label{eq:quench}
\end{align}
where at initial time, $\tau =0$, the values of  parameters are taken to correspond to those of the uniform wire at the sweet spot, (\ref{eq:HHazzard}): $\mu_b (0) = \mu = -t$, $t_b(0) = t, U_b (0) = U = 2t$, while at the end of the transformation, $\tau = \tau_f$,  $H_{\text{prep}} (\tau_f) = H_L + H_R + H_{\text{inter}} + 2 \mu N$. The term $2\mu N$ is a constant and added for convenience. 
%It removes local chemical potential at all sites except the edges of the wire where there remains an additional contribution $\mu n_1 + \mu n_L$, with $\mu = -t$. This edge chemical potential differential is necessary to ensure global SU(2) symmetry in the Jordan-Wigner related spin model which underpins the number degeneracy observed in this model, as we discuss in Sec.~\ref{sec:Heis}.
%$---the combination of terms at fixed $i$ in the sum in $H_L, H_R$ realize the SU(2)-invariant Heisenberg coupling $\propto \vec{S}_i \cdot \vec{S}_{i+1} + \text{const.}$ for $\mu = -t, U = 2t$. 
In what follows, we set $t = 1$. 
%\ivar{I think this discussing of SU(2) at this point is bit confusing. Also, lets drop the 2muN term so we need not explain it, and just note that mu on the edge sites is different to ensure degeneracy?}

Instead of performing a fully time-dependent simulation for an adiabatic transition between the initial and final Hamiltonians, for simplicity we instead consider a sudden quench, with $\tau_f \rightarrow 0$, and show that the protocol stably transitions from the ground state of the uniform wire to that of the partially split wire, despite the suddenness of the process. A truly adiabtic transition should excite the system less, and a lack of considerable heating after a sudden quench bolsters the claim that state preparation can be performed robustly. The quench corresponds to setting $\mu_b (0^+) = 0, U_b (0^+) = 0$ and $t_b (0^+) = t_{\text{inter}}$. 
%
%We consider two protocols for preparing $\Psi_g$. In the first case, values of $\mu_b, t_b, U_b$ are scaled synchronously and, at the end of the protocol, $\tau = \tau_f$, $\mu_b (\tau_f) = \mu \cdot t_{\text{inter}}/t , U_b (\tau_f) = U \cdot t_{\text{inter}}/t, t_b (\tau_f) = t_\text{inter}$. 
%In the second protocol, $\mu_b (\tau_f) = U_b (\tau_f) = 0$ while $t_b (\tau_f) = t_{\text{inter}}$. 
In Fig.~\ref{fig:quench} (a), we plot the results for a sudden quench  with $t_{\text{inter}} = 10^{-3}$ (note $t = 1$ as well as $U = 2, \mu = -1$). The wave function right after the quench  is still the ground state of the uniform wire, and  subsequently evolves under the dynamics of the quenched Hamiltonian $H_{\text{prep}} (\tau_f)$. While this represents a fairly violent quench, we find that the ground states before and after the  quench are  essentially the same.

%is obtained with a many-body overlap $\left|\langle \Psi (t) | \Psi_g \rangle \right|^2 \gtrsim 0.999$. 
%Thus, preparation of the vacuum state of the disjoint wires proceeds with a similar level of robustness as in the mean field setting. %\ivar{We should say that right after the quench the wavefunction is the same as it was before}

%\ivar{at the beginning we talk about adiabatic prep; but then we move into sudden quench. Explain why? that it is nice we need not do things adiabatically, a bonus? }

The reason for this robustness lies in the fact that the ground state of the uniform wire Hamiltonian is also the ground state of $H = H_L + H_R$ (see \ref{sec:wire-subwire}), also Sec.~\ref{sec:Heis} where we explain this result by invoking effective SU(2) symmetry in the related Jordan-Wigner transformed spin model. In fact, any SU(2) symmetric perturbation (satisfying the same relation between $U_b, \mu_b, t_b$ as in the uniform wire) does not affect the system at all. We confirm this in Fig.~\ref{fig:quench} (b); to numerical accuracy, we observe no excitation of the system whatsoever within the ground state manifold or beyond it. 
%One can most easily understand this in the spin language -- so weakened bond does not break the SU(2) symmetry of the Heisenberg model equivalent to (\ref{eq:HHazzard}); see Sec.~\ref{sec:Heis}.
%
%One can better understand this by appealing to the Jordan-Wigner transformed spin-$1/2$ model which is nothing but the Heinseberg spin chain. The ground state is fully polarized with total spin $S = L/2$. To construct such a state, any pair of spins-$1/2$s must fuse into a spin-$1$ state in order for the entirety of the spin chain to yield a total spin $S = L/2$. Thus, $\vec{S}_i \cdot \vec{S}_j$ for any $i,j$ leaves the ground state unchanged and yields an eigenvalue $\left[ 1(1+1) - 2 \times 1/2(1/2+1) \right]/2 = 1/4$. 
%

The quench we study can thus be viewed as a perturbation in the local chemical potential and interaction strength of magnitude $t_{\text{inter}}$. Given that $t_{\text{inter}} \ll t$, this quench generically will result only in excitation within the ground state manifold which disperses above the ground state over an energy window of $\mathcal{O} \left( t_{\text{inter}} \right)$. This is confirmed in the top figure (the projection to the ground state manifold is denoted by the operator $\text{P}_G$) in Fig.~\ref{fig:quench} (a) where we see that the system remains within the ground state manifold to accuracy greater than $0.999$, or $\sim 1 - t_{\text{inter}}/t$. However, it is not a small quench within the ground state manifold itself, and we do see changes in the population of states at long times; see bottom figure of Fig.~\ref{fig:quench} (a). As we will show in the next section, matrix elements of such local operators as the chemical potential and local interaction terms decrease exponentially in the eigenstate index within the ground state manifold, and also the system size $L$. Thus, we expect state preparation to be quite robust provided large changes respect SU(2) symmetry in the corresponding spin Hamiltonian.  
%\ivar{this very good. Just the last sentence, can make it more epxlicit what you mean?}

%It  breaks the $SU(2)$ symmetry in the corresponding spin Hamiltonian. Thus, we do see some excitation in the second case, as shown in Fig.~\ref{fig:quench} (b) where the system remains to good accuracy in the ground state subspace (top figure; the projection to the ground state manifold is described by the operator $\text{P}_G$), but over time, other states in that subspace can get populated. 

\begin{figure}
    \includegraphics[width=0.5\textwidth]{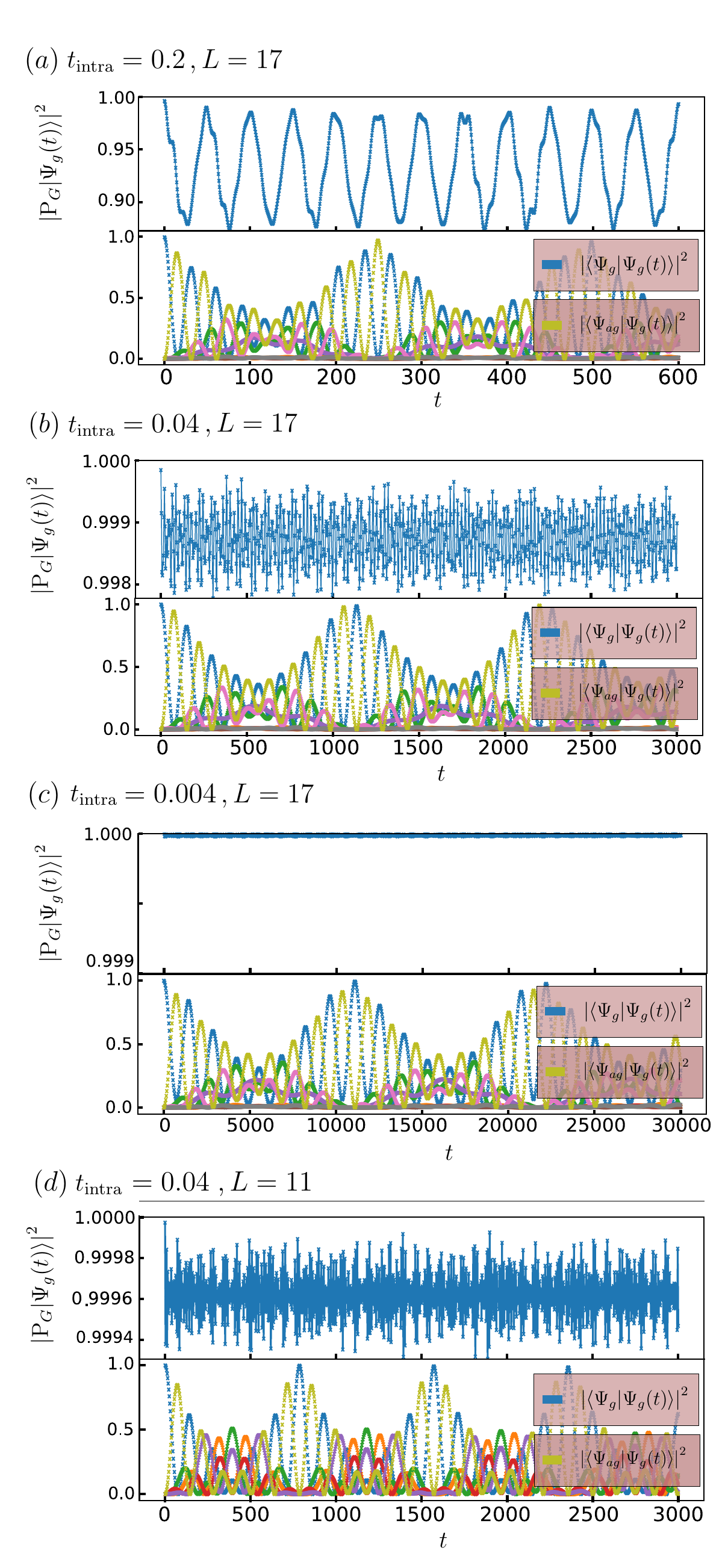}
    \caption{(a)-(d) Transition from $\Psi_g$ to $\Psi_{ag}$ using Hamiltonian braiding by introducing an intrawire coupling of amplitude $t_{\text{intra}}$ connecting left and right ends of the left split wire, for different values of $t_{\text{intra}}$ and system size $L$ (noted in figure); an extra plot for even larger $t_{\text{intra}} = 0.8$ is included in Fig.~\ref{fig:tintrahigh}. Top plot in each subplot shows projection of the time-evolved state to the ground state subspace of weakly coupled wires, while the bottom plot shows the projection to specific states in the ground state manifold, highlighting in particular the ground and anti-ground states, $\Psi_g, \Psi_{ag}$ respectively.}
    \label{fig:revivals}
\end{figure}

\subsection{Robustness in Braiding}

Next, we investigate numerically braiding using the Hamiltonian approach outlined in Section \ref{sec:HamiBraid}. Specifically, after preparing the ground state $\Psi_g$,  we introduce a hopping matrix element connecting the first and last sites of the left subwire. In the mean-field case, such a coupling would generate an effective low-energy term $i t_{\text{intra}} \gamma_{1} \gamma_{2}$ composed of the edge MZMs associated with the left subwire. After time $\tau= \pi/2t_{\text{intra}}$, as noted above, we expect to make the transition $\Psi_g \rightarrow - i \Psi_{ag}$. To simplify computations, instead of turning on such a coupling adiabatically, we again consider instead an abrupt quench with

\begin{align}
H (\tau < 0) &= H_\text{prep} (\tau_f) \equiv H_L + H_R + H_{\text{inter}},\\
H (\tau \ge 0) &= H_{\text{prep}} (\tau_f) + H_{\text{intra}}. 
\end{align}

\begin{figure}
    \includegraphics[width=0.4\textwidth]{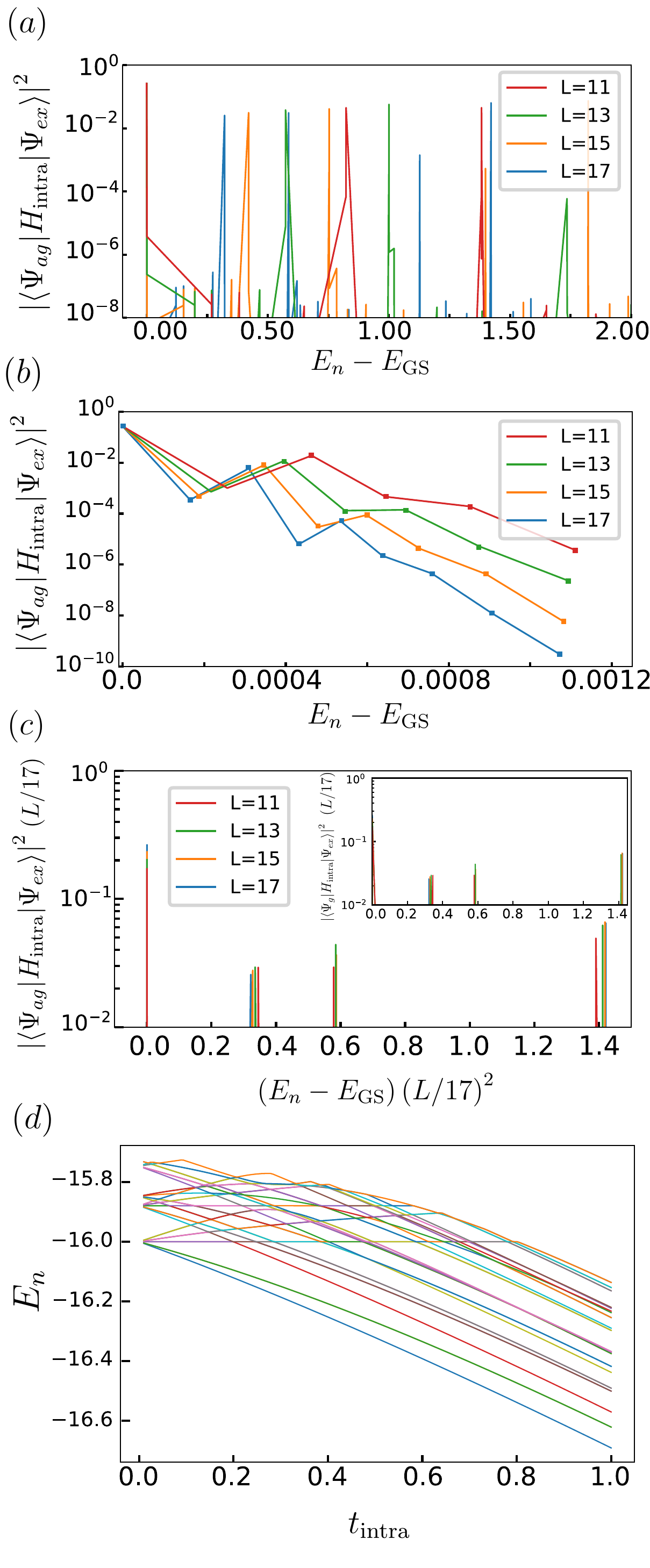}
    \caption{(a) Matrix elements of the intra-wire coupling $H_{\text{intra}}$ used to perform Hamiltonian braiding between $\Psi_{ag}$ and other states. (b) Close up of data in (a) shown restricting attention to matrix elements of $\Psi_{ag}$ with other states in the ground state manifold. Crucially, the matrix element $\matrixel{\Psi_g}{H_{\text{intra}}}{\Psi_{ag}}$ is independent of system siz. (c) Scaling collapse of data in (a) for matrix elements of $\Psi_{ag}$ with states outside the ground state manifold. (d) Evolution of the eigenspectrum due to intrawire coupling $t_{\text{intra}} = 0$ (no coupling) to $t_{\text{intra}} = 1$. There is a conspicuous lack of avoided level crossings visible on the scale of many-body gaps illustrating the weakness of these couplings.}
    \label{fig:matelements}
\end{figure}
Here $H (\tau = 0^{-}), H (\tau = 0^+)$ are the system Hamiltonians before and after the quench at $\tau =0$, and the system is prepared in the ground state of two weakly coupled subwires, $H (\tau = 0^{-})$, that is $\Psi_{g}$. We consider the overlap of the time evolved system wavefunction $\Psi_g (\tau)$ after the quench with states in the ground state subspace of $H (\tau = 0^{-})$, especially focusing on the overlap of $\Psi_g (\tau)$ with states $\Psi_g$ and $\Psi_{ag}$. We also compute the total amplitude of $\Psi_g(\tau)$ projected on to the complete ground state subspace and denote it via $\left| \text{P}_G \ket{\Psi_g (\tau)} \right|^2$. 

The main numerical findings are illustrated in Fig.~\ref{fig:revivals}, and Fig.~\ref{fig:tintrahigh}. We find that the state of the system always evolves from $\Psi_g$ to $\Psi_{ag}$ with a peak amplitude that appears to improve with increasing system size, for a broad range of coupling strengths $t_{\text{inter}} = 10^{-3} \lesssim t_{\text{intra}} \lesssim \mathcal{O}(1)$; see Figs.~\ref{fig:revivals} (a)-(d) for quenches with different $t_{\text{intra}}$ and system size $L$ (see also Fig.~\ref{fig:tintrahigh}) . The lower end of this range is simply set by the inter-wire coupling as is natural---for couplings $t_{\text{intra}} < t_{\text{inter}}$, $t_{\text{intra}}$ is a weak perturbation which does not significantly change the spectrum and the eigenstates of the coupled two-wires system.
The upper end of this range, on the other hand,  is larger than the naive expectations based on the excitation gap would suggest (which is about $0.1$ for $L=17$). 
Note that we are looking at the overlap of the $\Psi_g (t)$ on the ground state manifold of the Hamiltonian that does not include $H_{\text{intra}}$. As $H_{\text{intra}}$ grows non-perturbatively large, it becomes important potentially to consider the overlap with states in the ground state manifold at finite $H_{\text{intra}}$, which will be higher than the results plotted---the presence of revivals in the data is  suggestive of this possibility. 
Thus, it appears that $t_{\text{intra}}$ can be very large and still drive robustly transitions between $\Psi_g$ and $\Psi_{ag}$, which is remarkable in the many-body strongly interacting setting. As an aside, we note the presence of beating phenomena in the amplitudes within the ground state subspace with a characteristic frequency $\sim 1/L$, determined by the energy splitting within this  subspace. 

To better understand the robustness of the Hamiltonian braiding protocol, we examine the matrix elements of $H_{\text{intra}}$ connecting $\Psi_{ag}$ to other eigenstates of $H (\tau = 0^{-})$, that is, $\left| \matrixel{\Psi_{ag}}{H_{\text{intra}}}{\Psi_{ex}} \right|^2$. These results are shown in Fig.~\ref{fig:matelements} (a)-(c). First, within the ground state subspace, Fig.~\ref{fig:matelements} (b), the matrix element  $\matrixel{\Psi_{ag}}{H_{\text{intra}}}{\Psi_g}$ is finite and independent of system size while all other matrix elements appear to decrease exponentially with  the eigenstate index within the ground state subspace and with the system size. One can understand the above by noting that the action of $H_{\text{intra}} \ket{n} \propto {2 p (1-p)} (-1)^n \ket{n}$ when projected on the ground state subspace. The phase $(-1)^n$ arises because the tunneling of a particle by $H_{\text{intra}}$ from the first to the last site of the subwire results in a phase of $(-1)^{n-1}$ owing to fermionic statistics. The factor ${2 p (1-p)}$, where $p$ is the filling fraction of the $L_1$ subwire (see Appendix \ref{app:tun}), accounts for the number of states in $\Psi_{ag}$ which have have a particle on site $1$ and lack a particle on site $L_1$ and vice versa. This factor nominally depends on $n$, but since $|c_g^n|$, the weight of $\Psi_{ag}$ on $\ket{n}$, is peaked at $n = p L_1$ with a width $\delta n \sim 1/\sqrt{L}$, we can ignore these variations in the thermodynamic limit. Thus, $\matrixel{\Psi_g}{H_\text{intra}}{\Psi_{ag}} \approx {2p (1-p)} \sum_n |c^g_n|^2 = {2p (1-p)}$, i.e., remains finite in the thermodynamic limit. Note we have also used here the fact that the weights of $\Psi_g, \Psi_{ag}$ on the states $\ket{n}$ are given by $c_g^n, c_g^n (-1)^n$, respectively, and are tied to one another by an effective particle hole symmetry under which $H_{\text{inter}} \rightarrow - H_{\text{inter}}$ for $\ket{n} \rightarrow (-1)^n \ket{n}$. This symmetry is guaranteed provided $H_{\text{inter}}$ only involves single particle tunneling terms at the microscopic level (i.e., the Josepshon tunneling of Cooper pairs can be ignored).

%Additionally, this phase is precisely the difference between the wavefunctions $\Psi_g, \Psi_{ag}$. 

We now turn to the matrix elements of $H_{\text{intra}}$ between $\Psi_{ag}$ and other states in the ground state manifold, besides $\Psi_{g}$. As discussed in Appendix \ref{app:wf}, the states near the ground state and antiground states to good approximation are quantum harmonic oscillator states. They are a product of a Gaussian and the Hermite polynomial. 
For instance,  the $m$-th state $\Psi_{m}$ above the ground state $\Psi_{g}$ has $c_n^m \sim e^{-(n-n_0)^2/2\sigma^2 } H_m[(n-n_0)/\sigma] \propto c_n^{g}  H_m[(n-n_0)/\sigma]$. Near the peak, the Hermite polynomial generates oscillations with the wave vector $\propto \sqrt {2m}$. We can therefore approximate the matrix element between  $\Psi_{ag}$ and $\Psi_{m}$ as $\int dn \; |c_n^{g}|^2 \cos (\sqrt{2m} n/\sigma +\pi m/2)\sim e^{-m} $ for even $m$ and zero for odd $m$. Indeed, this is qualitatively the pattern that we see in Fig. \ref{fig:matelements}. Note for the largest $m \sim L$, one also obtains the exponential decrease with system size as seen in the data. For fixed small $m$, for instance $m = 2$, the data show decrease in the matrix element with system size $L$, but the decrease appears to be sub-exponential.

%For instance, $x = 0$ when evaluating the matrix element of $H_{\text{intra}}$ between $\Psi_{ag}$ and $\Psi_g$. This follows from the $n$-dependent phase structure of eigenstates in the ground state manifold; see Secs.~\ref{sec:wire-subwire} and Appendix~\ref{app:approx}. This integral amounts to evaluating the spectral weight of $f (q)$ at $q \approx x \pi$ which scales to zero exponentially both in system size $L$ and the eigenstate index, in agreement with numerical data. 
%The finiteness of the matrix element between $\Psi_g$ and $\Psi_{ag}$ on the other hand is protected by an effective particle hole symmetry $\ket{n} \rightarrow (-1)^n \ket{n}$ under which $H_{\text{intra}} \rightarrow - H_{\text{intra}}$ which dictates that the most excited state in the ground state manifold should have a sign structure $(-1)^n$ relative to the overall ground state. This further clarifies why both $\Psi_g, \Psi_{ag}$ are special within this ground state manifold. \ivar{lets try to stream line this. Why the matrix element given by that integral? why fn peaked at the typical number of particles? Why particle-hole matters?}

\begin{figure}
    \includegraphics[width=0.5\textwidth]{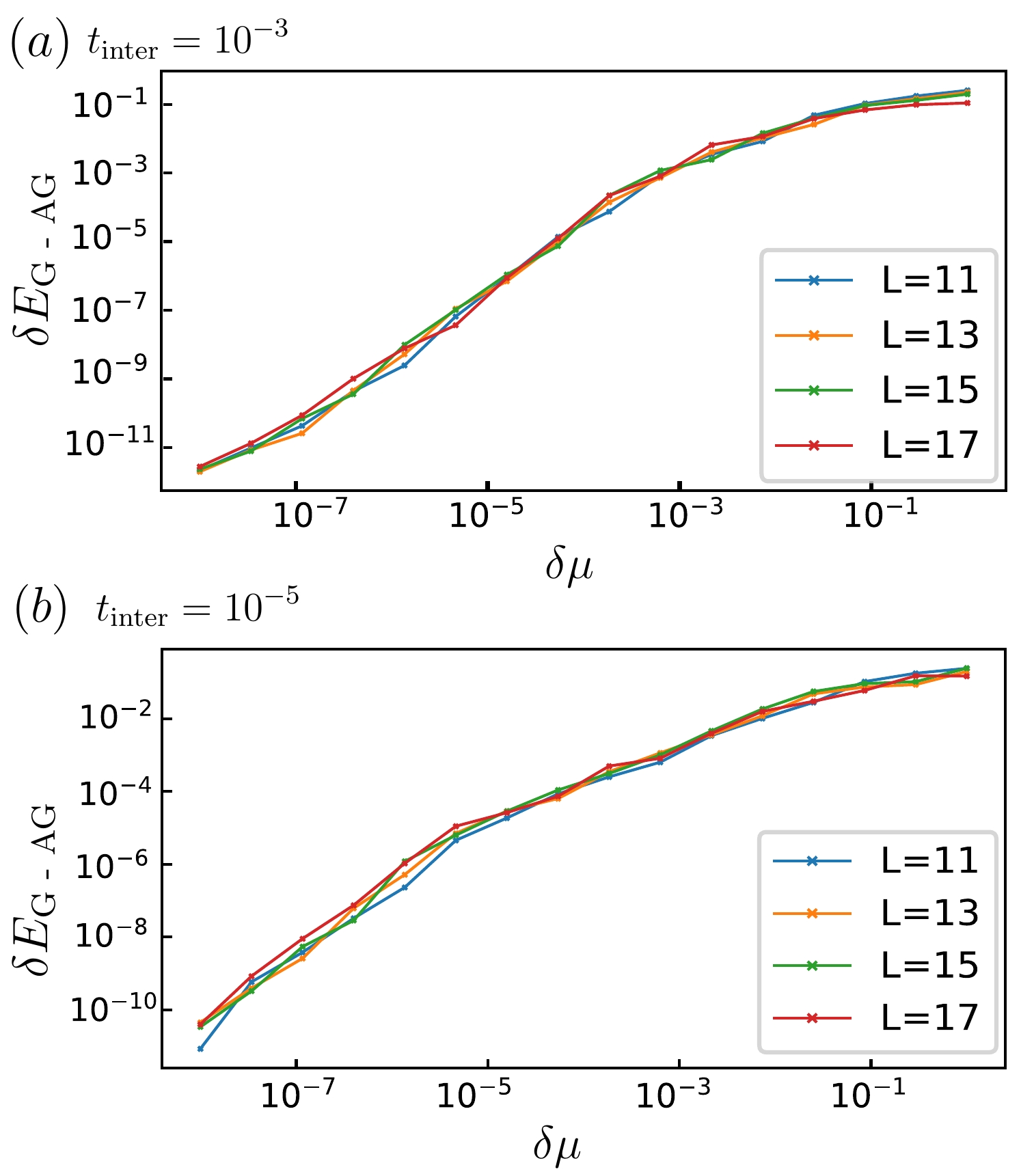}
    \caption{(a) Sensitivity of the energy difference between the ground and anti-ground states due to application of randomized chemical potential perturbations on each site of amplitude $\delta \mu$ (Gaussian distribution with zero mean and variance $\delta \mu^2$, for $t_{\text{inter}} = 10^{-3}$ and (b) $t_{\text{inter}} = 10^{-5}$. }
    \label{fig:dephasing}
\end{figure}

The above rationale provides a strong argument for why the transition from $\Psi_g$ to $\Psi_{ag}$ can proceed robustly provided $t_{\text{intra}}$ is smaller than the many-body gap. However, we observe clear transitions and coherent revivals for $t_{\text{intra}} \sim \mathcal{O} (1)$. To understand this, we examine the squared matrix elements $\left| \matrixel{\Psi_{ag}}{H_{\text{intra}}}{\Psi_{ex}} \right|^2$ connecting the anti-ground state $\Psi_{ag}$ to excited states $\Psi_{ex}$ outside of the ground state manifold. Remarkably, we find that despite the spectrum having a $1/L^2$ bulk gap between the ground state subspace and the rest of the states, there are only a  few matrix elements that are smaller than $\left| \matrixel{\Psi_{ag}}{H_{\text{intra}}}{\Psi_g} \right|^2$ by at most an order of magnitude; see Fig.~\ref{fig:matelements} (c). These squared matrix elements appear to decrease with system size at least as fast as $1/L$ while the energy of the corresponding state $\Psi_{ex}$ scales as $1/L^2$. Similar features are seen when considering matrix elements connecting the ground state $\Psi_g$ (as opposed to the anti-ground state discussed above) to  excited states; see inset of Fig.~\ref{fig:matelements} (c). 

We believe this sparsity and relatively small size of even the largest matrix elements connecting to other excited states is responsible for the robustness in transitions from $\Psi_g$ to $\Psi_{ag}$ despite the quasi-macroscopic degeneracy in the system. Visually, the weakness of the matrix elements can be seen in the evolution of the spectrum as a function of intra-wire coupling $t_{\text{intra}}$--- avoided level crossings of the state $\Psi_{ag}$ with excited states appear to be remarkably small on the scale of the gaps to excited states; see Fig.~\ref{fig:matelements} (d). In Sec.~\ref{sec:Heis}, we identify these spectral peaks with magnon modes of the equivalent Heisenberg model but are unable to explain the sparsity of the matrix elements. 
%\textcolor{red}{We unfortunately do not have a clear explanation for the sparsity of these matrix elements.}

Having established that $H_{\text{intra}}$ in its given form can reliably drive transitions between $\Psi_g$ and $\Psi_{ag}$, it is important to ask about the role of perturbations to $H^{\text{intra}}$. For instance, consider local perturbations to the chemical potential, $\delta V = \sum_i \delta \mu_i n_i$ . These operators do not connect states $\ket{n}, \ket{n'}$ for $ n \neq n'$. The matrix element $\matrixel{\Psi_{ag}}{\delta V}{\Psi_g} = \sum_n \delta V_n|c_g^n|^2 (-1)^n $ is then exponentially small in system size provided the matrix elements $\delta V_n \equiv \matrixel{n}{\delta V}{n}$ vary smoothly with $n$, similarly to $c_g^n$,  due to the fast-oscillating sign factors $(-1)^n$ distinguishing $\Psi_{ag}$ and $ \Psi_{g}$. In particular, we expect $\delta V_n \approx (n/L) \sum_i \delta \mu_i$ due to the fact that the states $\ket{n}$ involve equal superpositions of all possible distribution of particles on the subwires; thus the $\delta V_n$ do vary smoothly with $n$, completing our argument.

We also note that $H_{\text{intra}}$ needs not be  limited to tunneling between the first and last sites of the subwire. For instance, consider a tunneling perturbation $\delta H_{-1}$ between the first site of the left subwire and the second last site of the same subwire. $\delta H_{-1}$ does not  connect $\ket{n}, \ket{n'} $ for $n \neq n'$. Now, $\delta H_{-1} \ket{n}$ results in a state where a fraction $p = N/L$ of the states in $\ket{n}$ acquire a phase of $(-1)^{n-2}$ while a fraction $1-p$ acquire a phase $(-1)^{n-1}$. Thus, $\left|\matrixel{n}{\delta H_{-1}}{n} / \matrixel{n}{H_{\text{intra}}}{n}\right| \approx 1 - 2 p$ which vanishes around half filling, $p = 1/2$. Similarly, hopping between the first site to the third last site can be shown to be smaller than the corresponding matrix element of $H_{\text{intra}}$ by the factor $(1-2p)^2$, and so on. Thus, generically, longer range hopping terms are suppressed, analogously to the mean-field case. Moreover, all of them combined still effect the same transformation between $\Psi_{ag}$ and $ \Psi_{g}$ states, without causing any additional decoherence.

%the extra phases acquired by non-local tunneling operators will reduce the matrix elements exponentially in the distance from the edges of the subwire. This is entirely analogous to the mean-field setting wherein matrix elements connecting the topologically degenerate ground state subspace are decay exponentially away from the location of the MZMs at the ends of the (sub) wires. %This exponential decay is present in matrix elements of $\Delta H_{-1}, \delta H_{-2}, ...$ between other states in the ground state manifold as well---the robustness of matrix elements between $\Psi_g$ and $\Psi_{ag}$ to other states in the ground state manifold is important for robust realization of braiding protocols 
%\ivar{moral fixed?}

Finally, if we consider operators connecting the two wires, we note that these sign factors originating from the Fermion statistics single out two operators as having sizeable matrix elements connecting $\Psi_g$ and $\Psi_{ag}$---a hopping term connecting the left ends of the two subwires and the other connecting the right ends of the two subwires. These terms are also important in the usual mean field setting, corresponding to the coupling (or braiding) between Majorana operators $\gamma_1$ and $\gamma_3$, and $\gamma_2$ and $\gamma_4$, respectively (see Fig. \ref{fig:schematic}). Both drive the transition from $\Psi_g$ to $\Psi_{ag}$ (parity states of $i\gamma_2\gamma_3$). Thus, despite the additional macroscopic degeneracy in the ground state subspace of the interacting model, the matrix elements of physical operators are similar to those in the mean field case, showing how the topological robustness manifests  in the number conserving case. 

\subsection{Dephasing}

%First, let us note that here the operators with large diagonal matrix elements in the space of states $\ket{n}$ are local operators, by the reasoning given above. (Note the difference vis a vis operators with a large matrix element connecting the anti-ground and ground states which necessarily are ones connecting the ends of the subwires.)

We now discuss relative dephasing of the states $\Psi_g$ and $\Psi_{ag}$ by perturbations $\delta H$ for which $\delta E = \matrixel{\Psi_{ag}}{\delta H}{\Psi_{ag}} - \matrixel{\Psi_{g}}{\delta H}{\Psi_g} \neq 0$. Since $\Psi_g$ and $\Psi_{ag}$ are superpositions of number states with the same coefficients, one can already guess that there is likely to be some robustness to chemical potential fluctuations.  Consider again then the effect of $\delta V = \sum_i \delta \mu_i n_i$, a sum of local variations in the chemical potential. For such a perturbation, $\delta E = \matrixel{\Psi_g}{\delta V}{\Psi_g} - \matrixel{\Psi_{ag}}{\delta V}{\Psi_{ag}} = \sum_n \delta V_n \left( |c^n_g|^2 - |c^n_{ag}|^2 \right)$. The effective particle hole symmetry of $H_{\text{inter}}$, $\ket{n} \rightarrow (-1)^n \ket{n} \Rightarrow H_{\text{inter}} \rightarrow - H_{\text{inter}}$, implies that the coefficients $|c^n_g| = |c^n_{ag}|$ and $\delta E$ vanishes. Thus, such local terms do not cause dephasing between $\Psi_g, \Psi_{ag}$ at $\mathcal{O} (\delta \mu)$---that is, to first order in the perturbation, the states $\Psi_{ag}, \Psi_{g}$ can drift in energy but their relative energy difference remains zero. This is guaranteed as long as $H_{\text{inter}}$ satisfies the above effective particle hole symmetry which demands that only single particle hopping is allowed between subwires at the microscopic level. We verify this expectation numerically and present results in Fig.~\ref{fig:dephasing}. It is only for $\delta \mu$ larger than $t_{\text{inter}}$, at which point we cannot treat $\delta V$ as a perturbation, when we find $\delta E \sim |\delta \mu|$. We note that in this way, the protection from dephasing in weaker than in the standard mean field setting where all local terms result in an exponentially small correction (in the length of the subwire) to the splitting of degenerate ground states and at any order in a perturbative expansion. Here, although the first order correction $\propto \delta \mu$ appears to vanish as long as the above symmetry is enforced, a second order correction of magnitude $(\delta \mu)^2 / t_{\text{inter}}$ can persist.  

The above arguments extend to all generic local operators acting between states within the same subwire. Local operators connecting the two subwires, such as $H_{\text{inter}}$ of course do change the energy gap between these states but this is to be expected in the mean field setting as well.

\section{Mapping to magnetic systems}
\label{sec:Heis}

In this section we will exploit the correspondence between the fermionic and spin Hamiltonians to clarify some of the surprising features of the Hamiltonian (\ref{eq:HHazzard}) and its relation to Majorana fermions. First, as a point of reference, let us consider the mean-field Kitaev Hamiltonian. At the sweet spot $t = \Delta$, the transformation (see Sec.~\ref{sec:Heispre})  yields
\begin{eqnarray}
   H_{MF} &=& -\sum_{j = 1}^{L - 1} \{  a_{j}^{\dg}a_{j + 1}  - a_{j}a_{j + 1} + h.c. + \mu a_{j}^{\dg}a_{j}\} \\
   &\to& - \sum_{i = 1}^{L-1} \sigma^x_{i}\sigma^x_{i+1} - \mu  \sigma^z_{i}.
\end{eqnarray}
For $|\mu| \ll 1$, the model has two nearly degenerate states, separated from the rest of the spectrum by the $O(1)$ gap. They are $\ket{e} = (\ket{\uparrow} + \ket{\downarrow})/\sqrt{2}$ and $\ket{o} = (\ket{\uparrow} - \ket{\downarrow})/\sqrt{2}$
(here up and down refer to $all$ spins pointing in the $x$ or the opposite direction. In the fermion language, these states are the two Majorana parity ground states. 

Let's assume now that we weaken the bond between the sites $L_1$ and $L_1 + 1$, leaving only the single-electron hopping across it. We can define the two-wire basis states as $\ket{1s, 2s'}$, with $s$ and $s'$ being the global up and down states in the $x$ basis in the subwires 1 and 2, respectively.
The the single-electron tunneling matrix element is 
\begin{eqnarray}
    H_{\text{inter}} &=& - t_{\text{inter}} (a^\dag_{L_1}a_{L_1+1} +  a^\dag_{L_1+1} a_{L_1}) \nonumber\\
    &\to&t_{\text{inter}}(\sigma^+_{L_1}\sigma^z_{L_1}\sigma^-_{L_1+1}- \sigma^-_{L_1}\sigma^z_{L_1}\sigma^+_{L_1+1} ) \nonumber\\
    &=& -t_{\text{inter}}(\sigma^+_{L_1}\sigma^-_{L_1+1}+ \sigma^-_{L_1}\sigma^+_{L_1+1}). \label{eq:Ht1}
\end{eqnarray}
When acting in the ground states subspace we only need to retain $\sigma^x_{L_1}\sigma^x_{L_1+1}$; the eigenvalues are $+1$ for states $\ket{1s, 2s}$ and $-1$ for $\ket{1s, 2\bar s}$. Therefore if we start with the state $\ket{1e, 2e} = (1/2)\sum_s \ket{1s, 2s} + \ket{1s,2\bar s}$, 
the dynamical phase winds in the opposite directions for the first and the second terms. In particular, after certain time, the state will become $ i \sum_s \ket{1s, 2s} - \ket{1s,2\bar s} = i \ket{1o, 2o}$. So, the state does precisely what is expected from the Majorana picture after a double exchange of the edge Majoranas of the two subwires.

Now let's turn to the number-conserving model (\ref{eq:HHazzard}, \ref{eq:quench}) and its mapping to spin representation. The Jordan-Wigner transformation yields 
\begin{eqnarray}
   H_{} &=& -\sum_{i = 1}^{L - 1} \Big[ a_{i}^{\dg}a_{i + 1} + a_{i+1}^{\dg}a_{i} +\mu n_j + \mu n_{i+1} \nonumber \\ && + U n_i n_{i+1} \Big] \nonumber \\
   \to& - &\sum_{i = 1}^{L-1} \bigg[ (\sigma^x_{i}\sigma^x_{i+1} + \sigma^y_{i}\sigma^y_{i+1})/2 + \mu (\sigma^z_i + 1)/2 \nonumber \\
   &+& \mu(\sigma^z_{i+1} + 1)/2+ U  (\sigma^z_{i} +1) (\sigma^z_{i+1}+1)/4  \bigg].
\end{eqnarray}

As noted above, it has the sweet-spot Kitaev wave function as its ground state for $U = 2, \mu = -1$, where it corresponds to the isotropic ferromagnetic Heisenberg model. The ground state in this case is the fully-polarized ferromagnet. It is $L+1$-fold degenerate, which correspond to the states $\ket{S, m}$ with all possible values of $m=\sum_i S^z_i = \sum_i \sigma^z_i/2$, or equivalently, all possible fermion numbers, $n = L/2 + m$. The degeneracy between the ground states is lifted quadratically in $m$ when $U\ne 2$, and linearly in $m$ by finite fermion chemical potential $\mu$. An increase in the magnitude of $U$ favors easy axis ferromagnetism, which in the fermion language favors the $N= 0$ and $N = L$ states. In contrast, a decrease in the magnitude of $U$ favors easy-plane magnetism, with the lowest energy states near $m = 0$ or, equivalently,  $N = L/2$. 

Now let's split one long wire into two, of  $L_1$ and $L - L_1$ sites each. By themselves, they will have ground states that correspond to smaller multiplets with the total spin $S_1 = L_1/2$ in the first wire, and $S_2 = (L-L_1)/2$ in the second wire, yielding the basis set $\Psi = \ket{S_1, m_1; S_2, m_2}$. The tunneling (\ref{eq:Ht1})  commutes with the combined $S^z = m_1 + m_2$ of the wires (equivalent to the total fermion number conservation). In the ground-state subspace it therefore connects the states of the form
\begin{eqnarray}
     \ket{S_1, m; S_2, S^z - m} \equiv \ket{\psi_m}.\label{eq:subspace}
\end{eqnarray}
This is the spin counterpart of the wave function $\ket{N_1; N}$ discussed in section \ref{sec:2W}. The matrix elements $\bra{\psi_m}H_{\text{inter}}\ket{\psi_{m+1}}$ can be easily obtained from the spin algebra (Clebsch-Gordan coefficients); this is an alternative derivation of $t_n$  given in Appendix \ref{app:tun}. 
The derivation of the ground states in the spin language for  weak tunneling thus proceeds identically to the fermion language, with the ground and the antiground states having the form $   \Psi_g = \sum_{m} c_{m}^g \ket{\psi_m}$ and $ \Psi_{ag} = \sum_{m} (-1)^m c_{m}^g \ket{\psi_m}$. 
%These states correspond to the two junction Majorana parity states. 

We now are ready to contrast the special states $\Psi_{g/ag}$ and the parity states of the mean-field Kitaev wave function. The tunneling coupling (\ref{eq:Ht1}) favors parallel arrangement of the total spins in wires 1 and 2; however, all directions in the $x-y$ plane are now equivalent due to the $x-y$ symmetry of the Hamiltonian. Moreover, it is clear that for any linear combination of states (\ref{eq:subspace}), $\ket{\psi} = \sum_m a_m \ket{\psi_m}$ , $\bra{\psi} S_{1,2}^{x,y}\ket{\psi} = 0.$ On the other hand, $\bra{\Psi_{g,ag}} S_{1}^{x}S_{2}^{x} + S_{1}^{y}S_{2}^{y}\ket{\Psi_{g,ag}} \sim \pm S_1S_2$; that is, 
the ground and antiground states correspond to the correlated states where the spins of the two subwires are either aligned or anti-aligned in the $x-y$ plane, but without net magnetic moment in either subwire. This can be compared with the mean-field Kitaev (Ising) case discussed at the beginning of this section: there, the ground state (zero junction Majorana parity) was the superposition of subwire spins pointing in the same direction ($x$ or $-x$) for both wires, or the opposite direction in the case of antiground state.
In the number-conserving (Heisenberg) case, the  ground state is an equal superposition of both subwire spins co-aligned in any direction in the $x-y$ plane, or antialigned in the antiground state case.

One can now understand the robustness of the ground state wave function when a bond term $H_{\text{inter}}(\tau)$  is varied in strength, set by $t_b$, while maintaining the relationship between the the individual hopping, chemical potential, and interaction strengths $U_b =  2 t_b, \mu_b = - t_b$ [described in Eq. (\ref{eq:quench}) and illustrated in Fig. \ref{fig:quench}a]. The ground state of the whole system is fully polarized with total spin $S = L/2$. To construct such a state, any pair of spins-$1/2$s must fuse into a spin-$1$ state in order for the entirety of the spin chain to yield a total spin $S = L/2$. Thus, $\vec{S}_i \cdot \vec{S}_j = [(\vec{S}_i +\vec{S}_j)^2 - \vec{S}_i^2 -\vec{S}_j^2]/2$ for any $i,j$ leaves the ground state unchanged and yields an eigenvalue $\left[ 1(1+1) - 2 \times 1/2(1/2+1) \right]/2 = 1/4$. By maintaining the relationship between individual terms in $H_{\text{inter}} (\tau)$ to preserve the SU(2) symmetric form of the whole Hamiltonian, we guarantee that such an operator acts trivially on the ground state wave function for all times $\tau$.

Notably, the mapping from the fermion Hamiltonian yields a local spin Hamiltonian; also the tunneling between wires gives a local exchange interaction. The locality of the mappings is however violated by the intra-wire tunneling between the sites 1 and $L_1$, which we used above (section \ref{sec:HamiBraid}) to implement the equivalent of the intra-wire Majorana braiding. This operator is 

\begin{eqnarray}\label{eq:HTS2}
    H_{\text{intra}} &=& - t_{\text{intra}} (a^\dag_{L_1}a_{1} +  a^\dag_{1} a_{L_1}) \\
    &\to&-t_{\text{intra}}(\sigma^-_{1}\sigma^+_{L_1} + \sigma^+_{1}\sigma^-_{L_1}) \prod_2^{L_1 -1}(- \sigma_i^z)\\
    &=& -t_{\text{intra}}(\sigma^-_{1}\sigma^+_{L_1} + \sigma^+_{1}\sigma^-_{L_1})  P, \label{eq:Ht2}
\end{eqnarray}
where $P = \prod_1^{L_1}(- \sigma_i^z)$. The intrawire tunneling acts diagonally in the subspace (\ref{eq:subspace})$, \bra{\psi_m}H_{\text{intra}}\ket{\psi_{m}}\propto (-1)^{L_1/2 + m}$, alternating in sign between adjacent $m$ ($S^z$ states of the first wire). 
This is equivalent to the result in the fermion number basis (section \ref{sec:HamiBraid}), since $L_1/2 + m$ counts the number of fermions in the first wire.
In terms of the subwire macro-spins, the string of Pauli-$z$ operators acts to reverse the orientation of the first subwire spins, thus converting between the parallel and antiparallel relative orientations of spins in two subwires. This is the origin of the oscillation between the ground and the antiground states.

The mapping to the spins helps to shed some light on our observation that while implementing the Hamiltonian braiding, we can apply rather strong intrawire tunneling $t_{\text{intra}}$, comparable or larger than the spectral gap between the ground state manifold and the excited states; yet the excitation outside of the ground state manifold is very limited. This can be attributed to the suppression of hybridization between $\Psi_{ag}$ and the rest of the spectrum. 
To understand this suppression, let us examine which states are accessible from the anti-ground state by the action of $H_{\text{intra}}$. Combining the first and the last spins of the first subwire into a spin-1, the ground state of the $L_1$ wire can be represented as
\beq
\ket{S_1, m_1} = \sum_{m = -1,0,1} g_m \ket{S_1 - 1, m_1 -m; 1, m},
\eeq
where $g_m$ are the Clebsch-Gordan coefficients. The action of $H_{\text{intra}}$ annihilates all terms  expect  $m = 0$. On the other hand, in the limit of large $S_1$ and $m_1$, $\ket{S_1 - 1, m_1; 1, 0} = \sqrt{(1-z^2)/2}(\ket{S_1, m_1}- \ket{S_1 - 2, m_1}) + z\ket{S_1 - 1, m_1}$, where $z = m_1/S_1$. 
The states $\ket{S_1 - 1, m_1}$ and $\ket{S_1 - 2, m_1}$ contain one and two magnons, respectively. Note that typical $z$, which comprise the majority of the weight in $\Psi_{g/ag}$, are small, $z\sim 1/\sqrt{S_1}$.

Microscopically, application of $H_{\text{intra}}$ to $\Psi_{ag}$ can be thought of as conversion to $\Psi_{g}$ by string of Pauli-$z$, followed by the one- or two-magnon excitation due to the spin-flip of two edge spins of the first subwire, $H_{\text{intra}} \Psi_{ag}\sim \sigma_1^+\sigma_{L_1}^-\Psi_{g}$
A single magnon creation operator goes as $M_q^\dag \sim (1/\sqrt{L})\sum_j e^{i q r_j} \sigma_j^+$. We can therefore estimate matrix element of $H_{\text{intra}}$ between the antiground state and the ground state with one and two excited magnons as $t_{\text{intra}}z/\sqrt{L_1}$ and $t_{\text{intra}}/L_1$, respectively. Both scale approximately as $t_\text{intra}/L_1$, decreasing with increasing $L_1$. This is the rough estimate of the anticrossing gap between the antiground state and low-lying states. In practice, the avoided crossing gaps appear to be distributed unevenly, with most of them being strongly suppressed, and only a few gaps being sizable. 
This hints at the additional protection of the $\Psi_{g,ag}$ subspace, details of which we do not yet fully understand.

\section{Discussion}\label{sec:discussion}

The goal of this work was to understand how the braiding of Majorana zero modes, which in the mean field setting leads to nontrivial transformations in the ground state subspace, translates into the many-body setting beyond mean field. In addition to being theoretically interesting, this can have also practical implications as even subtle deviations from the idealized mean-field results may temper the appeal of topological quantum computing schemes that rely on such manipulations.

In the BCS mean field, the breaking of $U(1)$ charge conservation symmetry leads to the eigenstates that are superpositions of different fermion number states. For an isolated system this is clearly unphysical. 
In the number-conserving setting,  superconductivity  is characterized not by the global $U(1)$ symmetry breaking, but rather by a macroscopic degeneracy of ground states with different numbers of particles. In the conventional superconductors this translates into the ability to add the Cooper pairs to  the condensate at zero  cost, making all even parity states equal energy. In topological superconductors, $all$, both even and odd fermion number states become degenerate. 
This additional odd-even degeneracy in the mean-field treatment is a result of the appearance of the MZM's at the edges of the system.

We used number-projected Kitaev mean-field wave functions  as a specific example to gain understanding of the many body states of a topological superconductor.  These wave functions are the ground states of a spinless interacting fermions in a 1D chain \cite{Iemini15,WangNumberConserving}. 
For any fixed number of fermions, the ground state of such a wire is unique. However, splitting the wire into two subwires necessarily leads to a degeneracy, with any allocation of fermions between two subwires having the same ground state energy. It is this degeneracy that allows the establishment of the effective phase coherence between subwires and enables nontrivial manipulations within ground state subspace that is analogous to the Majorana braiding in the mean field. 

Indeed,  the two-wire wave function can be expanded in the  basis conjugate to the number of electrons in one of the wires, 
\beq
\ket{\varphi; N} = \sum_n e^{i n \varphi} \ket{n; N}
\eeq
and conversely
\beq
\ket{n; N} = \frac 1{2\pi}\int d\varphi e^{-i n \varphi}\ket{\varphi; N}.  
\eeq
For instance, in this basis, the ground state obtained by adiabatically splitting two wires (\ref{eq:Psi0}) becomes 
\begin{equation}
    \Psi_g = \int d\varphi  \, c^g_{\varphi} \, \ket{\varphi; N}  , \label{eq:Psi0phi}
\end{equation}
with $c^g_{\varphi} = \sum_{n} e^{-i n \varphi} c_{n}^g/2\pi$. 
Since $c_n$ is generally distributed over $\sim \sqrt{L_1}$ number states, in the relative phase basis, the wave function is concentrated in the narrow window of $\delta \varphi \sim 1/\sqrt{L_1}$, which becomes arbitrarily well defined in the large system limit. For the ground state $\Psi_g $ the phase is centered at $\varphi = 0$; for the anti-ground state $\Psi_{ag} $ -- at $\varphi = \pi$. Theses states are the many-body analogs of even and odd MZM junction parity states. 

The macroscopic degeneracy of the ground state of two fully split wires implies however that the states that we associated with the MZM parity states are not unique; rather, they depend on the details of the splitting procedure. For instance, if the wire is being split by weakening of not just a single bond, but also several adjacent bonds, the resulting state will be different. It is likely that this ambiguity will make the analogs of the MZM braiding manipulations not robust, unlike the mean-field case. That is even before we introduce other complicating factors such as the charging energy that will make different $\ket{n; N}$ state components evolve at different energies, causing loss of coherence.

While the MZM braiding scheme of \cite{alicea2011non}
has been conceptually very influential, more recently other schemes have been proposed that take into account the need for a common ``ground" superconductor that is both a source of Cooper pairs and eliminates the charging energy differences between different MZM parity states \cite{lutchyn2018majorana}. These schemes are no longer reducible to 1D wire geometries; nor do they necessarily require the geometric exchange of MZM's to realize quantum operations.  Adding explicitly the ground superconductor to the system effectively extends the boundaries of the system, of what needs to be included in the complete -- charge conserving -- treatment. 
How resilient the new architectures are with respect to the many body considerations of the kind that we studied in this paper is an open question.

\begin{acknowledgments}

The authors acknowledge useful discussions with C. D.  Batista, J. V\"ayrynen, and L. Rokhinson. This work was supported by the US Department of Energy, Office of Science, Basic Energy Sciences, Materials Sciences and Engineering Division.
\end{acknowledgments}

\appendix

\section{Splitting the wave function of a uniform wire} \label{app:psiu}

A special feature of the mean-field Kitaev ground state wave function at the ``sweet spot" $t = \Delta$ and $\mu = 0$ of the Hamiltonian \eqref{eq:HMF} is that it can be expressed in terms of the ground state wave functions of any two subsections. This can be easily seen using the Majorana representation, $a_i = \xi_i + i\eta_i$.  The Hamiltonian that satisfies $t_i = \Delta_i$ can be written as $H = \sum_{i= 1}^{L-1} 2 t_i P_i$, where $P_i = i\eta_i\xi_{i+1}$ is the Majorana bond parity \cite{kitaev2001unpaired}. Because all $P_i$ commute with each other, the general eigenstate can be labeled by the eigenvalues of these operators. The values of $t_i$ only determine the energies of the eigenstates, but the set of the eigenstates is identical for any choice of $t_i$'s. In particular, the eigenstates of a system with $t_i = const$ (uniform) and with $t_{i\ne i_0} = const$ and $t_{i_0} = 0$ are identical. The uniform system has two  ground states distinguished by the edge parity operator $P_0 = i \eta_L \xi_1$. These states are also the ground states of the system with the eliminated bond $t_{i_0} = 0$, and therefore can be expressed in terms of the product of ground states of the disconnected subwires. Each subwire can have either odd or even parity ground state, with the combined parity given by $P_0$. For example, for total even parity 
\beq
\ket{{\Psi_{e}}}_L = \alpha \ket{{\Psi_{e}}}_{L_1}\otimes \ket{{\Psi_{e}}}_{L_2} + \beta \ket{{\Psi_{o}}}_{L_1}\otimes \ket{{\Psi_{o}}}_{L_2}
\label{eq:Psi_e12}
\eeq
where the functions $\ket{\Psi_{o/e}}$ have the form of Eq. \eqref{eq:Psi}. [Note that the coefficients $\alpha$ and $\beta$ ($\alpha^2 + \beta^2 = 1$) are not fixed by this construction.] Finally, applying the number projection to the lhs and the rhs yields a linear combination over products of number-projected ground states of the subwires, of the same kind as \eqref{eq:PsiN_12}.

\section{Wavefunction with general weak bond}
\label{app:wf}

In this Appendix we will construct the ground state wave function of the superconducting wire with a fixed total number of particles, which is almost split in two parts by a weak link. Unlike Appendix \ref{app:psiu}, we will not assume any special point of the Kitaev model. We will only assume that the single electron tunneling between subwires is weak, and each subwire has a set of degenerate ground states distinguished only by their number of electrons. The higher-energy states are not involved because we assume that the tunneling is very weak.

Our starting point is the tunneling Hamiltonian \eqref{eq:HT}. We will rewrite it here in a more compact form

\begin{equation}
    H_T =  -\sum_{n} t_{n+0.5} \ket{n+ 1}\bra{n}+ t_{n-0.5}\ket{n- 1}\bra{n}. \label{eq:HT2}
\end{equation}
where as before $n$ is the number of electrons in the first subwire, and $\ket{n}$  is the state that we denoted as $\ket{N_1; N}$ in the main text, with $N_1 = n$.

The eignestates of this Hamiltonian $\Psi = \sum c_n \ket{n}$ satisfy the vector Schr\"odinger equation
\begin{eqnarray}
   E c_n  &=&   -t_{n+0.5} c_{n+ 1}- t_{n-0.5}c_{n- 1}.\label{eq:SE}
\end{eqnarray}
By the Perron-Frobenius theorem, the ground state has all positive amplitudes $c_n$, as long as all $t_n$ are positive. There is a symmetry connecting states with positive and negative energies: for every state with amplitudes $\{c_n\}$ and energy $E$, there is a state with the amplitudes $\{(-1)^n c_n\}$ and energy $-E$. In particular, this relationship connects the ground  and the anti-ground states. As we show in the main text, these are the states that can be associated with the distinct junction Majorana parity states in the mean-field treatment.

Assuming that $t_{n}$ is a smooth function $n$,  we can take the continuum limit in the Schroedinger equation
\begin{eqnarray}
   E c_n 
   &\to& - t_n (c_{n+ 1}+ c_{n- 1}) - \frac{t'_n}{2}(c_{n+ 1}- c_{n- 1})\\
   &\to& - t c'' - {t'}c' -2t c\label{eq:SE2}
\end{eqnarray}
In the last line, both $t$ and $c$ stand for functions of  variable $n$ promoted to be continuous. Ignoring for the moment the second term that involves the first derivative $a'$, and expanding near the minimum $n_0$ of the potential $-2t(n) = -2t(n_0) + \alpha (n-n_0)^2/2$, we recognize the Schroedinger equation for a harmonic oscillator with  the mass $1/2 t(n_0)$ and the stiffness $\alpha = 2 t''(n_0)$. The ground state of such a harmonic oscillator is 
\beq
\Psi_0 \propto \sum_n e^{-(n-n_0)^2/2\sigma^2}\ket{n}, \label{eq:gs0}
\eeq
where $\sigma^2 = \sqrt{t(n_0)/t''(n_0)}$. The frequency of the oscillator is $\Omega = 2\sqrt{t''(n_0) t(n_0)}$. In App. \ref{app:tun} we show that the tunneling is a function of (intensive) filling rather than (extensive) electron number, vanishing at zero and full fillings. For fillings away from 0 and 1  we can estimate the curvature from $t''(n_0) L_1^2\sim t(n_0)$. That immediately implies that the width of the oscillator wave function  $\sigma \sim \sqrt{L_1}$ is much smaller than the expectation value of the electron number itself, $\bar N_1 = p L_1$ for the large $L_1\gg 1$ \footnote{Note that this result does not depend on the magnitude of tunneling, as long as it is small enough to allow expansion of the wave function in terms of the ground states of the subwires.}. 
This justifies the assumption of the constant mass in the Schroedinger equation that we made above. 
%On the other hand, the frequency of the small oscillations (equal to the level spacing of the oscillator states) can be estimated as $\Omega \sim t(n_0)/L_1$, within the same assumptions.

Now let's also verify that the term $t'a'$ could be ignored. Indeed, for the wave function that we just found, it can be estimated as $\alpha (n-n_0)^2 /\sigma^2$. This is much smaller than the potential term $\alpha (n-n_0)^2$ that we have already included in the derivation, and hence can be safely ignored.

Finally, in addition to the ground state Eq. (\ref{eq:gs0}), we can also construct the excited oscillator states. The energy spacing between these states is $\hbar \Omega \propto t(n_0)/L$. Therefore, the superpositions of these low energy states will evolve on the extensive time scale, $L/t(n_0)$. 
This timescale will provide a cut-off for the quantum coherence time for certain quantum gates based on splitting and fusing subwires that we discuss in the main text. This energy splitting also manifests in the beating pattern of revivals in the Hamiltonian braiding, Fig. \ref{fig:revivals}.

\section{Details of the tunneling Hamiltonian}
\label{app:tun}

Here we discuss the detailed form of the tunneling  Hamiltonian \eqref{eq:HT} to which the Hamiltonian \eqref{eq:HHazzard} reduces in the limit of a very weak tunneling bond connecting two subwires. 
Assuming that the tunneling matrix element connects only the last site of the first wire to the first site of the second wire, from Ref \cite{sajith2023signatures}, $t_{N_1} \propto [p_1(1-p_1) p_2 (1-p_2)]^{1/2}$, where $p_1 = N_1/L_1$ and $p_2 = N_2/L_2 = (N - N_1)/L_2$.  This result was obtained assuming that the ground state wave function is the number-projected Kitaev wave function, obtained at the ``sweet spot" of maximally localized MZM (hopping equal to the superconducting gap and zero chemical potential, which corresponds to the average filling $p = 0.5$).

For equal length subwires, $L_1 = L_2$, this  function is peaked at $N_1 = N_2$, regardless of the filling $p = N/L$.  For small deviations from the the peak value, $n = N_1 - p L_1$, we have
\bea
t_n \propto p(1-p) - \frac{p^2 + (1-p)^2}{2 p (1-p)}(n/L_1)^2   
\eea
Another important special case is $p = 0.5$, for which indeed the Kitaev wave function is the most appropriate. In this case, for any $L_1$ and $L_2$ the peak of $t_n$ occurs at $n = 0$:
\bea
t_n \propto 1 - 4  \left(\frac 1{L_1^2} + \frac 1{L_2^2}\right) n^2  .
\eea

On the other hand, if $L_1 \ll L$, then $t_{N_1}$ peaks at $L_1 /2$, which generally corresponds to a filling different from $N/L$, unless $p = 0.5$. That means that if the bond is weakened truly adiabatically,  the charge 
will flow in order to reach the value of $L_1/2$ in the short segment. In the presence of the charging energy, this flow will be opposed an additional potential that will tend to pin the subwire charge at the value of $pL_1$.  We discuss this in detail in Appendix \ref{app:charge}.

\section{Charging energy}
\label{app:charge}

As we  saw in Appendix \ref{app:wf}, the junction wavefunciton peaks at the value of $n_0$ (number of electrons in the subwire $L_1$) where $t_n$ reaches maximum. However, if the subwires are not of the same length or $p\ne 0.5$, this  does not correspond to the average filling of the original, long, wire. This implies that in the absence of charging energy, the densities of the subwires will become different if the splitting is done adiabatically. Of course, in a real system, the charging energy will counteract this effect as it will tend to keep the average electron density nearly constant, matched to the ionic density of the material. Let us consider the effect of the charging energy on the wave function. Again staying within the ground state subspace, the charging Hamiltonian can be written as 
\begin{equation}
    H_C =  \sum_{n} E_C(n - n^*)^2 \ket{n}\bra{n}. \label{eq:Hc}
\end{equation}
where $n^*$ corresponds to the charge neutrality. From the form of this Hamiltonian and comparison with Eq. (\ref{eq:HT2}), it is  clear that its role is to combine with the position dependent potential $-2t_n$ in order to redefine the equilibrium position of the effective harmonic oscillator (\ref{eq:SE2}). Generally, the presence of the charging energy will make the oscillator wave-function tighter, making the charge more localized than it would be only due to the curvature of the hopping $t_n$. In fact, as the tunneling goes to zero adiabatically, the width of the wave function disappears as $\sigma^2 = \sqrt{t_{n^*}/E_C}$, fully localizing charge and destroying the number superpositions that are created by wire splitting in the absence of the charging energy. Thus, to preserve superpositions, it is in fact necessary to break adiabaticiy 
while turning off the tunneling between the subwires, while the tunneling is still much larger than $E_C$. 

Note that due to the finite charging energy, even after the tunneling is turned off completely, the states with different $n$ will be evolving with different phases (due to their different energies, given by (\ref{eq:Hc})): If the width of the state at the moment of the loss of adiabaticity is $\sigma$, the time it will take for the different $n$ components of the wave function to decohere can be estimated as $T_\phi \sim  1/(E_C \sigma^2) = 1/\sqrt{t_d(n^*) E_c}$, where $t_{d}(n^*)$ is the value of tunneling at the time that adiabaticity is lost. This time happens to correspond to the Majorana version of the Josepshon plasmon period in a junction with constant tunneling $t_{d}(n^*)$. Given the above conditions on the presence of superposition, the decoherence rate is certainly larger than $E_C$.

\section{Other (approximate) wave functions in ground states subspace}

\label{app:approx} 

In the main text we focused on  the ground and the anti-ground states of the tunneling Hamiltonian. The ground  state is an equal-sign superposition of states with different distributions of fermions between the two sides of the junction; this is quite  similar to the mean-filed state that describes two superconductors coupled by tunneling, with a  relative zero phase between them.

The phase in superconductor plays the role of a quasimomentum on the (electron) number lattice. Starting from the ground state we can also define other quasimomentum states as
\begin{equation}
    \Psi_\phi = \sum_{n} c^g_{n} e^{in \varphi} \ket{n; N}. \label{eq:Psiphi}
\end{equation}
When the two subwires are completely disconnected, these states can be generated from the reference ground state $\{c_n^g\}$ by applying a potential difference pulse between  the two subwires.

For general $\varphi$ these states are not the  eigenstates of the tunneling Hamiltonian \eqref{eq:HT}. 
However, for $\varphi$ near 0 or $\pi$ they are sufficiently close to the harmonic oscillator regime that we discussed in App. \ref{app:wf} to be expandable in terms of the low-lying harmonic oscillator states. Therefore, their dynamics will be slow, on the time scale proportional to the system size. Within this timescale, we can treat them as approximate eigenstates, with the expectation value of the energy $E= -2t_{n_0} \cos\varphi$.

 The states (\ref{eq:Psiphi}) are the approximate eigenstates not only of Hamiltonian, but also of the current operator, $\hat I = i[H_T, N_1] $ with the eigenvalues $ I = 2t_{n_0}\sin \varphi$. In states $\Psi_0 = \Psi_{g}$ and $\Psi_\pi = \Psi_{ag}$ the current expectation value is zero, while it is non-zero for other $\varphi$. 

\section{Hamiltonian braiding with larger $t_{\text{intra}}$}
\label{sec:appextra}

\begin{figure}
    \includegraphics[width=0.5\textwidth]{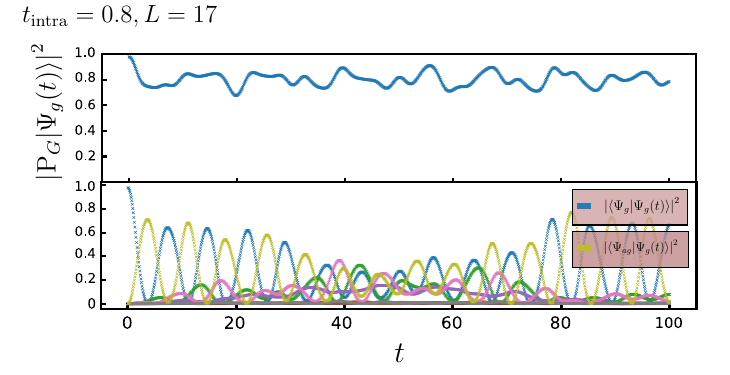}
    \caption{Hamiltonian braiding with $t_{\text{intra}} = 0.8$, system size $L = 17$, and $N = 8$ particles. Even for $t_{\text{intra}} \sim \mathcal{O} (1)$, and much larger than the many-body gap, we see fairly strong transitions between $\Psi_g, \Psi_{ag}$ and multiple revivals which strongly suggests that there is little leaking out of the ground state manifold. (Note that the projection to the ground state manifold in the top figure is computed based on the manifold at $t_{\text{intra}} = 0$.)}
    \label{fig:tintrahigh}
\end{figure}

We provide further data showing that Hamiltonian braiding can be performed robustly even for $t_{\text{intra}} \sim \mathcal{O} (1)$, Fig. \ref{fig:tintrahigh}. 

 %Also, these are the only states stabilized by weak interwire tunneling in the presence of even small changing energy -- the states with other $\phi$ would lead to current between the subwires, whose charging energy would prevent such states from being stationary. \textcolor{red}{need to consider cases of pi and 0, may be only one of them is stabilized by charging energy?}

%\bibliographystyle{apsrev4-2}
%\bibliography{refs}


\end{document}